\documentclass[11pt,a4paper]{article}

\usepackage{amssymb,amsmath}
\usepackage{graphicx,graphics}

\usepackage[english]{babel}
\usepackage{epsfig,url}
\usepackage{bbm,theorem}
\usepackage{a4wide}

\usepackage{amssymb}
\usepackage{amsmath}
\usepackage{amsfonts}
\usepackage{bbm}
\usepackage{latexsym}

\newcommand{\BE}{Boltzmann equation}

\newcommand{\FT}[1]{\widehat{#1}}
\newcommand{\Eharm}{H^{(0)}}
\newcommand{\Weql}{W^{(\text{eq})}}

\def\w#1{\mathop{:}\nolimits\!#1\!\mathop{:}\nolimits}

\newcommand{\Z}{{\mathbb Z}}
\newcommand{\R}{{\mathbb R}}
\newcommand{\C}{{\mathbb C\hspace{0.05 ex}}}
\newcommand{\vep}{\varepsilon}
\newcommand{\qand}{\quad\text{and}\quad}

\newcommand{\ombasic}{\overline{\omega}}

\newcommand{\T}{{\mathbb T}}
\newcommand{\1}{\mathbbm{1}}
\newcommand{\rme}{{\rm e}}
\newcommand{\ci}{{\rm i}}
\newcommand{\rmd}{{\rm d}}

\newcommand{\order}[1]{\mathcal{O}(#1)}
\newcommand{\braket}[2]{\langle #1 , #2\rangle}
\newcommand{\bigbraket}[2]{ \Big\langle #1 , #2\Big\rangle}
\newcommand{\mean}[1]{\langle #1\rangle}
\newcommand{\defem}[1]{{\em #1\/}} 

\newcommand{\set}[1]{\{#1\}}
\newcommand{\norm}[1]{\Vert #1\Vert}

\newcommand{\tr}{\operatorname{Tr}}
\newcommand{\re}{\operatorname{Re}}

\newcommand{\cf}{{\mathbbm 1}} 
\newcommand{\vc}[1]{\mathbf{#1}}

\newcounter{jlisti}

\DeclareMathOperator*{\sign}{sign}

\begin{document}

\newcommand{\email}[1]{E-mail: \tt #1}
\newcommand{\emailjani}{\email{jani.lukkarinen@helsinki.fi}}
\newcommand{\addressjani}{\em University of Helsinki,
Department of Mathematics and Statistics\\
\em P.O. Box 68,
FI-00014 Helsingin yliopisto, Finland}

\title{Kinetic theory of phonons in weakly anharmonic particle chains}
\author{Jani Lukkarinen\thanks{\emailjani}\\[1em]
$^*$\addressjani}

\date{\today}

\maketitle

\begin{abstract}
The aim of this review
is to develop the kinetic theory of phonons in classical particle chains
to a point which allows comparing the kinetic theory of normally conducting chains,
with an anharmonic pinning potential, to the kinetic theory of the anomalously conducting FPU chains.  
In addition to reviewing existing results from the literature, we present
a streamlined derivation of the phonon Boltzmann collision operators using Wick polynomials, as well as details about
the estimates which are needed to study the effect of the collision operator.  This includes explicit solutions of the collisional constraints,
both with and without harmonic pinning.  We also recall in detail the derivation of the Green--Kubo formula for thermal conductivity in these systems,
and the relation between entropy and the Boltzmann H-theorem for the phonon Boltzmann equations.  The focus is in systems which are spatially translation invariant
perturbations of thermal equilibrium states. We apply the results to obtain detailed predictions from kinetic theory for the Green--Kubo correlation functions,
and hence the thermal conductivities, of the chain with a quartic pinning potential as well as the standard FPU-$\beta$ chain.
\end{abstract}

\tableofcontents

\section{Kinetic scaling limit for weakly anharmonic chains}

\subsection{Introduction}

Kinetic theory describes motion which is \defem{transport dominated} in the sense that 
typically the solutions to the kinetic equations correspond to constant velocity, i.e., \defem{ballistic}, motion intercepted by 
\defem{collisions} whose frequency is order one on the kinetic space-time scales.
The constant velocity part of the transport, the \defem{free streaming motion}, 
can arise via several different microscopic mechanisms, the obvious case being free motion of classical particles. 
Here, we are interested in evolution of energy density in chains composed out of locally interacting  
classical particles, in the limit where the interactions are dominated by a stable harmonic potential.
For such models, the energy transport is dominated by the free streaming of \defem{phonons} which correspond to 
eigenmodes of the motion generated by the harmonic potential. 

Our goal is to derive and study the properties of kinetic theory given by a phonon Boltzmann equation.
Even with most generous estimates for the terms neglected in its derivation, 
the Boltzmann equation can become an exact description of the energy transport 
only in a \defem{kinetic scaling limit} which produces  
a total scale separation between the free streaming motion of phonons and the collisions induced by 
the perturbation.
For the above particle chains this amounts to studying lengths and time intervals both of which are proportional 
to $\lambda^{-2}$, where $\lambda$ is the strength of the anharmonic perturbation, and then taking
the limit $\lambda\to 0$.  

In particular in one dimension, i.e., for particle chains, it is difficult to estimate 
reliably the error terms associated with
the approximation of the original motion by the phonon Boltzmann equation.
At present, no such full mathematical analysis has been achieved.
However, there are instances in which 
the one-dimensional phonon Boltzmann equation has proven to be useful:
in Section \ref{sec:kappapinning}, the thermal conductivity predicted
by the Boltzmann equation of systems with an anharmonic \defem{onsite} potential
is compared to numerical simulations, and the results are found to agree
within numerical accuracy for sufficiently weak couplings.

Although just one example, the agreement is significant since it shows
that, under proper conditions, the kinetic theory from the phonon Boltzmann 
equation can produce meaningful predictions even for one-dimensional transport problems.
As will be apparent later, computing these predictions requires
some additional effort compared to the standard rarefied gas Boltzmann equation.
The effort is rewarded in a prediction which has \defem{no adjustable parameters}
and hence can be compared directly with experiments and numerical simulations.

The drawback of the phonon Boltzmann equation is that, \defem{a priori}, 
it only describes the lowest order effects of the perturbation.  Hence,
it might not capture all processes relevant to the energy transport.
In addition, since it is not known which precise conditions guarantee that
the corrections are truly vanishing in the kinetic scaling limit, 
one has to leave the option open that even the lowest order effects are
not being described accurately by the equation.

A discussion about the physical and mathematical conditions under which the phonon 
Boltzmann equation should be an accurate approximation is given in \cite{spohn05},
while \cite{Ziman67} could serve as a textbook reference on physics of phonons.
The present rigorous results on linear and nonlinear perturbations
of wave-like evolution equations support the basic principle that 
for sufficiently dispersive systems, in particular for crystals in three
or higher dimensions, the derivation of the phonon Boltzmann equation 
yields an accurate approximation to the behaviour of the 
system at kinetic time-scales with weak coupling.  The argument is particularly convincing
for spatially homogeneous systems and for linear perturbations such as
when the particle masses have a weak random disorder \cite{butz15,ls05},
and for the closely related quantum models, random Schr\"odinger equation 
\cite{erdyau05a,erdyau99} and the
Anderson model \cite{chen03}.
In addition, it has been proven that
the Wigner transforms of unperturbed wave-equations do, in great
generality, converge to the collisionless Boltzmann equation 
\cite{gerard97,mielke05}, although acoustic modes may require somewhat more
refined treatment \cite{hltt06}.  
Nonlinear perturbations are more intricate, and we refer to 
\cite{BCEP08,NLS09} for recent results on the rigorous analysis of their
kinetic scaling limits.

Quite often the addition of the collision operator to the transport equation leads 
to loss of ``memory'' at each collision and eventual equilibration via diffusion: 
this would correspond to normal energy transport by phonons.  
However, if the transport is anomalous, it is highly likely to be reflected also
in the solutions of the corresponding kinetic equation.  An example of such
behaviour is given by the FPU-$\beta$ chain.  The aim of this contribution
is to develop the general kinetic theory of classical particle chains
to the point which allows to compare the kinetic theory of 
normally conducting chains
with an anharmonic pinning potential and the anomalously conducting FPU chains.

The first section concerns the derivation of the phonon Boltzmann equation 
in the one-dimensional case, for general dispersion relations and
third and fourth order perturbations of the harmonic potential. 
We begin with the infinite volume harmonic model in Sec.~\ref{sec:freephonons}, 
and make a brief comment on the finite volume case
and the relevance of boundary conditions in Sec.~\ref{sec:phononBC}.
The anharmonic perturbations are discussed in Sec.~\ref{sec:anharmpert}
where we also explain our choice for the definition of the related energy density 
and derive the related energy current observables.

The role of the proper choice of initial data for the derivation 
of the Boltzmann equation cannot be stressed enough.  We describe our choices
in Sec.~\ref{sec:initialdata}, and introduce the notations needed in the
analysis of the cumulants of the phonon fields.  We conclude the first section
with a derivation of the Green--Kubo formula for the particle chains, under the assumption
that energy density satisfies a closed diffusion equation.  (This would
correspond to a case with only one locally conserved field.)

The first section concerns fairly general particle chains, but in order 
to derive the phonon Boltzmann equation, further restrictions are required.
In Sec.~\ref{sec:deriveC}, we do the derivation in detail for the FPU chains,
and sketch it for the onsite anharmonic perturbations.  The rest of the
section concerns only chains with stable \defem{nearest neighbour} harmonic couplings.
We explain why the nearest neighbour couplings do not produce any collisions
at the kinetic time-scale from the third order terms of the potential  (Sec.~\ref{sec:C3iszero}),
and show that only phonon number conserving collisions appear from the fourth order
terms (Sec.~\ref{sec:C4gen}).  To use the collision operator, one needs to 
resolve the related collisional constraints.  This is done for FPU-chains 
in Sec.~\ref{sec:FPUconst} and for the chains with a harmonic pinning potential 
in Sec.~\ref{sec:solOnsite}.

The final section is devoted to the analysis of the resulting
class of phonon Boltzmann equations.  We first show how an 
assumption of an increase of microscopic
entropy in homogeneous models leads to an H-theorem for the phonon \BE .
In Sec.~\ref{sec:BEsteadys}, it is explained how the H-theorem can be used to 
classify all steady states of the \BE .
The kinetic description derived in Sec.~\ref{sec:GKformula} for
the time correlation function in the Green--Kubo formula is applied to these models in Sec.~\ref{sec:linearizedBE}.
This results in explicit predictions for the leading asymptotics of the 
decay of the correlations.  The decay is found to be integrable 
for the onsite perturbation, and we make a comparison of the resulting 
thermal conductivities in Sec.~\ref{sec:kappapinning}.
We discuss the corresponding prediction in FPU models in Sec.~\ref{sec:CtFPU}.
The result is decay $O(t^{-3/5})$, which predicts anomalous conduction in the FPU chains.

More comments and discussion about the results are given in Sec.~\ref{sec:conclusions},
before the Acknowledgements and References.

\subsection{Free motion of phonons}\label{sec:freephonons}

Phonons correspond to eigenmodes of harmonic Bravais lattices.  
The general case includes multi-component phonon fields which may arise both 
from the dimensionality of the lattice---the evolution equation could 
describe \defem{displacements} from some reference lattice positions and thus have $d$ components in $d$ dimensions---or 
from the reference cell of periodicity containing 
more than one particle.  
An example of the latter is provided by the one-dimensional chain with alternating masses: then the reference system remains invariant only under shifts over two lattices sites which 
corresponds to a Bravais lattice with two particles per cell of periodicity.  
For notational simplicity, let us here only consider the case of one-component classical phonon fields and assume that all particles have a unit mass (this is always possible to achieve for one-component fields by choosing a suitable time-scale).

Then the state of the system at time $t$ is determined by $(q_x(t),p_x(t))$, $x\in \Z$, which satisfy the evolution equations
\begin{align}\label{eq:harmev}
 \dot{q}_x(t) = p_x(t),\quad \dot{p}_x(t) = -\sum_y \alpha(x-y) q_y(t)\, .
\end{align}
The equations are of a Hamiltonian form, corresponding to the Hamiltonian function 
\begin{align}
H(p,q)=\sum_{x\in\Z}\tfrac{1}{2}p^2_x+ 
\sum_{x,y\in\Z}\tfrac{1}{2} q_x \alpha(x-y) q_y\,. 
\end{align}
We suppose that the harmonic interactions have a \defem{short range}.  For instance, 
only finitely many $\alpha(z)$ are different from zero, or
$|\alpha(z)|$ decreases exponentially fast to zero as $|z|\to \infty$.  

The evolution equation (\ref{eq:harmev}) can be solved by using Fourier-transform.  We define the Fourier-fields as
$\FT{q}(t,k) = \sum_{x\in \Z} \rme^{-\ci 2\pi k x} q_x(t)$ and $\FT{p}(t,k) = \sum_{x\in \Z} \rme^{-\ci 2\pi k x} p_x(t)$.  Here 
$k\in \T$ where $\T$ denotes the one-torus which we parametrize using the interval $[-\tfrac{1}{2},\tfrac{1}{2}]$ and then identify its endpoints,
$-\tfrac{1}{2}$ and $\tfrac{1}{2}$.  The evolution equation for the Fourier-fields then reads, in a matrix form,
\begin{align}\label{eq:FTharmev}
 \frac{\rmd }{\rmd t}\begin{pmatrix}\FT{q}(t,k) \\ \FT{p}(t,k)
                     \end{pmatrix}
                     = \begin{pmatrix}
  0 & 1\\ -\FT{\alpha}(k) & 0\end{pmatrix} \begin{pmatrix}\FT{q}(t,k) \\ \FT{p}(t,k)
                     \end{pmatrix} \, .
\end{align}
For each $k$, the equation is easily solved by Jordan decomposition of the matrix on the right hand side.
This provides two independently
evolving eigenmodes $a_t(k,\sigma)$, $\sigma=\pm 1$, as long as $\FT{\alpha}(k)\ne 0$.   For those $k$ with $\FT{\alpha}(k)=0$, the solution 
is given by constant speed increase: $\FT{p}(t,k) = \FT{p}(0,k)$ and $\FT{q}(t,k) = \FT{q}(0,k)+ t \FT{p}(0,k)$.  Typically, this can happen at most at 
a finite number of points whose neighbourhoods might require special treatment in the kinetic theory of phonons.

The eigenmode fields $a_t(k,\sigma)$, which we call \defem{phonon modes} of the harmonic chain, satisfy the evolution equation
\begin{align}\label{eq:atevol}
\frac{\rmd }{\rmd t} a_t(k,\sigma)=-\ci \sigma \omega(k) a_t(k,\sigma) \, ,
\end{align}
where $\omega(k)=\sqrt{\FT{\alpha}(k)}$. 
Thus the square root of the Fourier-transform of $\alpha$ determines the \defem{dispersion relation} $\omega$ of the phonons,
and we arbitrarily choose the principal branch of the square root so that always $\re \omega(k)\ge 0$.
There are several possible ways to normalize the eigenmodes.  Here we employ the following choice, 
standard to phonon physics, 
\begin{align}\label{eq:deffta}
a_t(k,\sigma)=\frac{1}{\sqrt{2\omega(k)}}\left( \omega(k) \FT{q}(t,k)
 + \mathrm{i} \sigma  \FT{p}(t,k) \right), \quad \sigma \in \{\pm 1\},\ k\in
\T \, .
\end{align}

It is straightforward to check that these fields indeed satisfy the evolution equation (\ref{eq:atevol}) whenever $\omega(k)\ne 0$.  
In addition, if we choose a solution to (\ref{eq:atevol}) and then define
\begin{align}\label{eq:invphonon}
\FT{q}(t,k) = \frac{1}{\sqrt{2\omega(k)}} \sum_{\sigma}a_t(k,\sigma)\qand
\FT{p}(t,k) = \frac{\sqrt{2\omega(k)}}{2} \sum_{\sigma}(-\ci \sigma) a_t(k,\sigma)\, ,
\end{align}
then $(\FT{q}(t,k),\FT{p}(t,k))$ yields a solution to (\ref{eq:FTharmev}).  
Given some initial field $a_0$,
the solution of (\ref{eq:atevol}) is straightforward, yielding
$a_t(k,\sigma) = \rme^{-\ci \sigma \omega(k) t} a_0(k,\sigma)$.  Therefore, if we let $a_0$ to be determined by the initial 
data $(\FT{q}(0,k),\FT{p}(0,k))$, we can now obtain the solution to the original 
evolution problem (\ref{eq:FTharmev})  for every $k$ for which $\FT{\alpha}(k)\ne 0$, just by 
inserting $a_t(k,\sigma) = \rme^{-\ci \sigma \omega(k) t} a_0(k,\sigma)$ in (\ref{eq:invphonon}).

For later use, it is also important to assume that the harmonic interactions are \defem{stable}, i.e., they
do not have any solutions which increase exponentially in time.   
By the above discussion, this is equivalent to assuming that the Fourier-transform of $\alpha$ is pointwise non-negative,
$\FT{\alpha}(k)\ge 0$ for all $k\in \T$.
A standard example is given by the \defem{nearest neighbour interactions} for which $\alpha(z)=0$ if $|z|>1$.  A short computation reveals that the only stable
nearest neighbour dispersion relations are given by $\omega(k) =\ombasic (1-2\delta \cos(2\pi (k{-}k_0)))^{1/2}$ for some
$\ombasic>0$, $|\delta|\le \tfrac{1}{2}$ and $k_0\in \T$.  Since for phonons we also require that $\alpha(z)$ are real, we can also always choose $k_0=0$ above.

The parameter $\ombasic$ determines the average angular frequency of the oscillations, 
the maximum value begin $\ombasic \sqrt{1+2 |\delta|}$ and the minimum $\ombasic \sqrt{1-2 |\delta|}$.  If $|\delta|<\tfrac{1}{2}$, the function $\omega(k)$ is a thus an even, real analytic, function.  These systems describe \defem{optical phonon modes}, and one can also think of the system as having \defem{harmonic pinning}.
If $\delta<0$, it is possible to make a change of variables 
which reverses its sign to positive: this is achieved with a shift by $\tfrac{1}{2}$ in the Fourier variable,
in other words, by using $((-1)^x q_x,(-1)^x p_x)$ as the new lattice fields.  
Enforcing this convention has the benefit that then the slowest evolving fields are also spatially slowly varying, 
corresponding to small values of $k$.  Therefore, we only consider 
the cases $k_0=0$ and $0<\delta\le \tfrac{1}{2}$ in the following, excluding also the degenerate
constant dispersion relation corresponding to $\delta=0$.

If $\delta=\tfrac{1}{2}$ one has $\omega(k) =\ombasic \sqrt{2} |\sin(\pi k)|$, with an $|k|$-singularity near origin.  These systems describe \defem{acoustic phonon modes} and the 
seemingly innocuous singularity has important implications for the energy transport properties of the system.
The FPU-chains belong to this category.

Finally, let us point out that in the stable case, $\omega(k)\ge 0$, 
the phonon fields defined by (\ref{eq:deffta}) 
are directly connected to the energy of the harmonic system by 
$\frac{1}{2}\sum_\sigma \int\rmd k \omega(k) |a_t(k,\sigma)|^2= H(p(t),q(t))$.
Hence, in analogy to quantum mechanics, 
the complex phonon fields $a_t(k,\sigma)$ can be thought of as wave functions of
phonons each of which carries an energy $\omega(k)$.  

\subsubsection{Effect of boundary conditions in finite non-periodic chains}
\label{sec:phononBC}

For notational simplicity, the above discussion uses formally an infinite lattice setup.  
Its proper interpretation is to consider the infinite
system to be an approximation for a finite, but large system.  This identification has an important mathematical consequence: the standard methods of $\ell^p$-spaces do not immediately apply to the infinite system.  For instance, typical samples from infinite volume Gibbs states are only logarithmically bounded at infinity \cite{lll77}.  Hence, 
should the need arise, it is better to go back to the finite systems
to resolve any possible issues about existence of solutions or to handle singularities.  (For discussion about the existence and properties of the infinite
volume dynamics, see \cite{lll77,butta07}.)

In a finite system, the choice of boundary conditions begins to play a role.  The above computations involving Fourier-transforms can be given an exact correspondence by choosing a box with periodic boundary conditions.
In this case, the wave evolution can move energy around the lattice unimpeded, as indicated by the solutions 
which are given by multiplication operators in the Fourier-space.  One can find more details about the correspondence,
for instance, in \cite{NLS09}.

If the finite lattice does not have periodic boundary conditions, 
the boundaries will influence the transport of phonons.  For boundary conditions
which preserve energy, such as Dirichlet and Neumann, 
the effect can be understood as a reflection of phonons at the boundary.  
It is also possible to combine these with partial periodic transmission through the boundary.  

The effect of standard heat baths, such as Langevin heat baths, on the kinetic equations
is more difficult to analyse.  
These will introduce also absorption and source terms to the phonon evolution equations, 
but since the heat baths are typically acting only on a few particles of the chain, their coupling to the wave 
evolution is nontrivial and somewhat singular.  (In general, it is difficult 
to affect the long wavelength part of the evolution by any local change to the system.)
Unfortunately, a systematic analysis of these effects appears to be missing from the current literature.

\subsection{Particle chains with anharmonic perturbations}
\label{sec:anharmpert}

For the discussion about the effect of local nonlinear perturbations to the above wave equation, let us introduce, 
in addition to the harmonic coupling function $\alpha:\Z\to \R$, 
its second and third polynomial counterparts, $\alpha_3:\Z^2\to \R$ and $\alpha_4:\Z^3\to \R$.
These are assumed to be ``small'' in the sense that for large microscopic 
times the evolution is dominated by the linear term.  Explicitly, we take the perturbed evolution equations to be
\begin{align}\label{eq:anharmev}
 & \dot{q}_x(t) = p_x(t)\, ,\nonumber\\
 & \dot{p}_x(t) = -\sum_y \alpha(y) q_{x-y}(t) - \sum_{y_1,y_2} \alpha_3(y_1,y_2) q_{x-y_1}(t)q_{x-y_2}(t)
 \nonumber \\ & \qquad
 - \sum_{y_1,y_2,y_3} \alpha_4(y_1,y_2,y_3) q_{x-y_1}(t) q_{x-y_2}(t) q_{x-y_3}(t)\, .
\end{align}
This corresponds to Hamiltonian evolution with the interaction potential
\begin{align}\label{eq:defVtot}
 & \mathcal{V}(q) = \frac{1}{2}\sum_{x,y} \alpha(y) q_x q_{x-y} +\frac{1}{3} \sum_{x, y_1,y_2} \alpha_3(y_1,y_2) q_x q_{x-y_1}(t)q_{x-y_2}(t)
 \nonumber \\ & \qquad
 +\frac{1}{4} \sum_{x,y_1,y_2,y_3} \alpha_4(y_1,y_2,y_3) q_x q_{x-y_1}(t) q_{x-y_2}(t) q_{x-y_3}(t)\, ,
\end{align}
supposing, as we will do here, that the following symmetries hold:
\begin{align}%\label{eq:harmev}
 & \alpha(y) = \alpha(-y)\, ,\\
 & \alpha_3(y_1,y_2) = \alpha_3(y_2,y_1) =\alpha_3(-y_1,y_2-y_1) \, , \\
 & \alpha_4(y_1,y_2,y_3) = \alpha_4(y_2,y_1,y_3) =  \alpha_4(y_1,y_3,y_2)
 =\alpha_4(-y_1,y_2-y_1,y_3-y_1)
\, .
\end{align}
The symmetries arise from assuming that the coefficients
correspond to a generic label-exchange symmetric, translation invariant potential.
We have also used here the fact that any permutation can be expressed as a product of adjacent transpositions.
Therefore, it suffices to check symmetry
with respect to the transpositions  to get full permutation invariance.

For example, the standard FPU-chain is given by
\begin{align}%\label{eq:harmev}
 & \mathcal{V}_{\rm FPU}(q) = \sum_{x} U(q_x -q_{x-1}),
\quad
U(r) = \frac{1}{4} \ombasic^2 r^2 +\lambda_3\frac{1}{3} r^3 +\lambda_4\frac{1}{4} r^4\, ,
\end{align}
and it satisfies the evolution equation $\ddot{q}_x = U'(q_{x+1}-q_x)-U'(q_x-q_{x-1})$.  A brief computation,
using the periodicity, shows that the FPU chain corresponds to choosing the coefficient functions above so that 
their nonzero values are
\begin{align}%\label{eq:harmev}
  \alpha(y) & = \frac{1}{2} \ombasic^2\begin{cases}
                2, & y=0\\
                -1, & y \in \{ \pm 1\}
               \end{cases} ,\\
  \alpha_3(y_1,y_2) & = \lambda_3 \begin{cases}
                 1, & y_1=y_2=1,\ y_1=-1,y_2=0,\text{ or } y_1=0,y_2=-1\\
                -1 , & y_1=y_2=-1,\ y_1=1,y_2=0,\text{ or } y_1=0,y_2=1
               \end{cases} \, , \\
  \alpha_4(\mathbf{y}) & = \lambda_4 \begin{cases}
                2, & y_1=y_2=y_3=0\\
                 1, & \pm\mathbf{y}\in \{(0,1,1),(1,0,1),(1,1,0)\}, \\
                 -1, & \pm\mathbf{y}\in \{(1,1,1),(1,0,0),(0,1,0),(0,0,1)\}
               \end{cases} 
\, .
\end{align}
For later use, let us also record their Fourier-transforms.
Using the shorthand notations $p_i=2\pi k_i$, $i=1,2,3$, they
are given by
\begin{align}%\label{eq:harmev}
  \FT{\alpha}(k_1) & = \ombasic^2 (1-\cos p_1) = 2 \ombasic^2 \sin^2 \frac{p_1}{2} \, ,
 \\
  \FT{\alpha}_3(k_1,k_2) & 
 % = \ci \lambda_3 2 (\sin p_1 + \sin p_2 -\sin (p_1+p_2) ) \nonumber \\ & 
  =\ci \lambda_3 2^3 \sin \frac{p_1+p_2}{2} \sin \frac{p_1}{2}\sin \frac{p_2}{2} \, ,\\
  \FT{\alpha}_4(k_1,k_2,k_3) & = 
  \lambda_4 2 \re \Bigl[\prod_{\ell=1}^3 (1-\rme^{-\ci p_\ell})\Bigr]
 % \nonumber \\ & =  \lambda_4 2 \re \Bigl[(1-\rme^{-\ci \sum_\ell p_\ell})(1- \sum_{\ell=1}^3 \rme^{\ci p_\ell})\Bigr]
  %\nonumber \\ &
  = -\lambda_4 2^4 \sin \frac{p_1+p_2+p_3}{2} \prod_{\ell=1}^3 \sin \frac{p_\ell}{2}
\, .
\end{align}

Onsite perturbations also belong to the above category.  For instance, 
\begin{align}%\label{eq:harmev}
\mathcal{V}_{\rm OS}(q) = 
\sum_{x,y\in\Z}\tfrac{1}{2} q_x \alpha(x-y) q_y + \sum_{x} V(q_x),
\quad
V(q) = \lambda_3\frac{1}{3} q^3 +\lambda_4\frac{1}{4} q^4\, ,
\end{align}
has
\begin{align}%\label{eq:harmev}
 & \alpha_3(y_1,y_2) = \lambda_3\cf(y_1=y_2=0)\, , \\
 & \alpha_4(y_1,y_2,y_3) = \lambda_4\cf(y_1=y_2=y_3=0)
\, .
\end{align}
Both Fourier-transforms are thus constant: $\FT{\alpha}_3(k_1,k_2)= \lambda_3$ and $\FT{\alpha}_4(k_1,k_2,k_3)= \lambda_4$.

After these examples, let us come back to the general case, equations (\ref{eq:anharmev}).  
Taking a Fourier-transform yields
\begin{align}\label{eq:FTanharmev}
& \frac{\rmd }{\rmd t}\FT{q}(t,k) =\FT{p}(t,k) \, ,\nonumber\\
& \frac{\rmd }{\rmd t}\FT{p}(t,k) =-\FT{\alpha}(k)\FT{q}(t,k)
 - \int_{\T^2}\! \rmd k'_1 \rmd k'_2\, \delta_\T(k-k'_1-k'_2) \FT{\alpha}_3(k'_1,k'_2) \FT{q}(t,k'_1)\FT{q}(t,k'_2)
 \nonumber \\ & \qquad
 - \int_{\T^3}\! \rmd k'_1 \rmd k'_2\rmd k'_3\, \delta_\T(k-\sum_{\ell=1}^3 k'_\ell) 
  \FT{\alpha}_4(\vc{k'}) \prod_{\ell=1}^3 \FT{q}(t,k'_\ell)\, ,
\end{align}
where we have denoted the periodic $\delta$-function by $\delta_\T$.  We will drop the subscript in the following and merely use $\delta$.
This implies that the phonon fields, defined still by (\ref{eq:deffta}), now satisfy the evolution equation
\begin{align}\label{eq:atevolanh}
& \frac{\rmd }{\rmd t} a_t(k_0,\sigma_0)=
 -\ci \sigma_0 \omega(k_0) a_t(k_0,\sigma_0)
 \nonumber \\ & \qquad
 -\ci \sigma_0 \sum_{\sigma_1,\sigma_2\in \set{\pm 1}}
 \int_{\T^2}\! \rmd k_1 \rmd k_2\, \delta(k_0-k_1-k_2) 
 \Phi_3(k_0,k_1,k_2)  a_t(k_1,\sigma_1) a_t(k_2,\sigma_2)
 \nonumber \\ & \qquad
 -\ci \sigma_0 \sum_{\sigma_1,\sigma_2,\sigma_3\in \set{\pm 1}}
  \int_{\T^3}\! \rmd k_1 \rmd k_2\rmd k_3\, \delta(k_0-\sum_{\ell=1}^3 k_\ell) 
  \Phi_4(k_0,k_1,k_2,k_3)
 \prod_{\ell=1}^3 a_t(k_\ell,\sigma_\ell)
 \, ,
\end{align}
where the interaction amplitude functions are 
\begin{align}\label{eq:defAmp}
 \Phi_3(k_0,k_1,k_2) & =  \FT{\alpha}_3(k_1,k_2)
   \prod_{\ell=0}^2 \frac{1}{(2\omega(k_\ell))^{\frac{1}{2}}}\, , \\
 \Phi_4(k_0,k_1,k_2,k_3)& =
  \FT{\alpha}_4(k_1,k_2,k_3)   \prod_{\ell=0}^3 \frac{1}{(2\omega(k_\ell))^{\frac{1}{2}}}  \, .
\end{align}
Since the $\delta$-function enforces the sum of the integration variables to be equal to $k_0$, we 
can  thus employ the following definitions for the FPU-chains 
\begin{align}\label{eq:defPhiFPU}
 \Phi_3(k_0,k_1,k_2) & = \ci\lambda_3 2^{\frac{3}{4}} \ombasic^{-\frac{3}{2}} \prod_{\ell=0}^2 g(k_\ell)\, , \\
 \Phi_4(k_0,k_1,k_2,k_3)& = -2 \lambda_4 \ombasic^{-2} \prod_{\ell=0}^3 g(k_\ell)\, , \label{eq:defPhiFPU4}
\end{align}
where  $g(k) = \sign(k) \sqrt{|\sin(\pi k)|}$ for $|k|\le \frac{1}{2}$.

To study energy transport, it is necessary to split the total energy into local
components.  The harmonic energy, corresponding to the energy of free phonons, is most conveniently
distributed using a form which is symmetric in Fourier-space.  By the definition (\ref{eq:deffta}),
we have for any $k_1,k_2$
\begin{align}
 \sqrt{\omega(k_1)\omega(k_2)} \sum_{\sigma=\pm 1} a(k_1,-\sigma) a(k_2,\sigma) = 
 \omega(k_1)\omega(k_2) \FT{q}(k_1) \FT{q}(k_2)
 + \FT{p}(k_1) \FT{p}(k_2) \,.
\end{align}
Thus, if we denote the inverse Fourier transform of $\omega$ by $\widetilde{\omega}$,
we can define
\begin{align}
& \Eharm_x(p,q) =
 \frac{1}{2} p_x^2 + 
\frac{1}{2}\left(\sum_{y} \widetilde{\omega}(y) q_{x-y} \right)^2
% \nonumber \\ & \quad
 = 
 \frac{1}{2} \sum_{\sigma=\pm 1} \int_{\T^2}\! \rmd k \,\rme^{\ci 2 \pi  x (k_1+k_2)} 
 \Phi_2(k) a(k_1,-\sigma) a(k_2,\sigma)\, ,
\end{align} 
where $\Phi_2(k_1,k_2)=\sqrt{\omega(k_1)\omega(k_2)}$.  Since $\omega$ is real and symmetric, we have here $\widetilde{\omega}(x)\in \R$.
Thus $\Eharm_x\ge 0$ and 
$\sum_x \Eharm_x(p,q)$ is equal to the harmonic part of the total energy, to
$\frac{1}{2} \sum_x  p_x^2+\frac{1}{2}\sum_{x,y}\tfrac{1}{2} q_x \alpha(x-y) q_y$.
Thus this choice allows distributing the positive total harmonic energy into positive contributions
localized at each lattice site.

For the anharmonic terms of the potential energy 
we use the form suggested by the notation in (\ref{eq:defVtot}), and 
define the local energy at site $x$ by 
\begin{align}\label{eq:phononendens}
 & H_x(p,q)
 = \Eharm_x(p,q) +\frac{1}{3} \sum_{y_1,y_2} \alpha_3(y_1,y_2) q_x q_{x-y_1}q_{x-y_2}
 \nonumber \\ & \qquad
 +\frac{1}{4} \sum_{y_1,y_2,y_3} \alpha_4(y_1,y_2,y_3) q_x q_{x-y_1} q_{x-y_2} q_{x-y_3}
 \nonumber \\ & \quad
 = 
 \frac{1}{2} \sum_{\sigma=\pm 1} \int_{\T^2}\! \rmd k \,\rme^{\ci 2 \pi  x (k_1+k_2)} \Phi_2(k) a(k_1,-\sigma) a(k_2,\sigma)
  \nonumber \\ & \qquad
 + \frac{1}{3}\sum_{\sigma\in \set{\pm 1}^3}
 \int_{\T^3}\! \rmd k\, \rme^{\ci 2 \pi  x \sum_{\ell=1}^3 k_\ell} 
 \Phi_3(k) \prod_{\ell=1}^3 a(k_\ell,\sigma_\ell)
  \nonumber \\ & \qquad
 + \frac{1}{4}\sum_{\sigma\in \set{\pm 1}^4}
 \int_{\T^4}\! \rmd k\,\rme^{\ci 2 \pi  x \sum_{\ell=1}^4 k_\ell} 
 \Phi_4(k) \prod_{\ell=1}^4 a(k_\ell,\sigma_\ell)\, .
\end{align}
Then clearly $\sum_x H_x(p,q) = \frac{1}{2} \sum_x  p_x^2 +  \mathcal{V}(q) = H(p,q)$, and  
and $H_x$ is a local function at $x$ (it depends mainly on fields near the point $x$). 
In addition, $H_x$ is translated just like the fields if these are shifted by $x_0$: if we choose some $x_0\in \Z$ and define 
$\tilde{q}_x=q_{x+x_0}$ and $\tilde{p}_x=p_{x+x_0}$, then $H_{x}(\tilde{q},\tilde{p})=H_{x+x_0}(q,p)$.

There are various possible choices for how to split the total energy into local energy density.
The above definition of energy density is perhaps not the most standard one, but it has two appealing features
for those systems which are dominated by phonon transport.  First, it has simple algebraic dependence on the phonon eigenmode fields.  Secondly, its dependence on the position variable $x$ is located entirely in the Fourier-factor.  This allows 
a definition of a current observable with a simple dependence on the particle interactions and the simplest 
choice for the associate discrete derivative.  

Explicitly, consider some $k\ne 0$ and any $x_1,x_2\in \Z$ with $x_1\le x_2$.
Setting $y=x_2-x_1$, we then have $y\ge 0$ and
\begin{align}
 & \sum_{x=x_1}^{x_2}\rme^{\ci 2 \pi  x k} =
 \rme^{\ci 2 \pi  x_1 k} \sum_{y'=0}^{y} \rme^{\ci 2 \pi  y' k}
 = 
 \rme^{\ci 2 \pi  x_1 k} \frac{1- \rme^{\ci 2 \pi (y+1) k}}{1- \rme^{\ci 2 \pi k}}
%  \nonumber \\ & \quad 
 = \frac{\ci \rme^{-\ci \pi k}}{2 \sin(\pi k)}
 \left[\rme^{\ci 2 \pi  x_1 k}-\rme^{\ci 2 \pi  (x_2+1) k}\right]
 \, .
\end{align}
Hence, if we define $J_{x-1,x}(t)$ by replacing in (\ref{eq:phononendens})
every factor 
``$\rme^{\ci 2 \pi  x \bar{k}} \prod_{\ell} a(k_\ell,\sigma_\ell)$'', where $\bar{k}=\sum_{\ell} k_\ell$,
by ``$\frac{\ci \rme^{-\ci \pi \bar{k}}}{2 \sin(\pi \bar{k})} \rme^{\ci 2 \pi  x \bar{k}} \partial_t\left( \prod_{\ell} a_t(k_\ell,\sigma_\ell)\right)$'', we have for all 
$x_1,x_2\in \Z$ with $x_1\le x_2$ that
\begin{align}\label{eq:latticecont}
 \partial_t\left( \sum_{x=x_1}^{x_2} H_x(q(t),p(t))\right) = J_{x_1-1,x_1}(t)-J_{x_2,x_2+1}(t)\, .
\end{align}
Since $x_1,x_2$ are arbitrary, we can interpret $J_{x,x+1}(t)$ as 
the energy flux from site $x$ to $x+1$ at time $t$.

To be precise, the above construction can only be used inside the integral if $\bar{k}\ne 0$.  To see why the resulting
current observable still works, let us for the moment regularize the system by 
replacing the infinite lattice $\Z$ by a finite periodic lattice of length $L\gg 1$.  Then $k_\ell \in \Z/L$, and
either $\bar{k}=\sum_\ell k_\ell\in \Z$ or the distance of $\bar{k}$ from $\Z$ is at least $1/L$.
Let us thus separate all terms with $\bar{k}\in \Z$ in the definition of $H_x$, and use the periodic identification.
Then the terms correspond to $\bar{k}=0$.  Since for every $x$ on the periodic lattice
$\rme^{\ci 2 \pi  x\bar{k}}\cf(\bar{k}{=}0) = \frac{1}{L}\sum_y \rme^{\ci 2\pi y \bar{k}}$, these terms sum 
to $H(p,q)/L$, i.e., to the total energy density.  By conservation of energy,
these terms do not contribute to $\partial_t H_x(q(t),p(t))$.  The remaining terms converge to the 
claimed result as $L\to \infty$, provided that the integrals over $\T^n$ are understood as principal value integrals
around the subset with $\bar{k}=0$.

In summary, the definition 
\begin{align}\label{eq:defencurrent}
 & J_{x-1,x}
 =\frac{1}{2} \sum_\sigma \int_{\T^2}\! \rmd k \, \Phi_2(k)
 \left.\frac{\ci \rme^{-\ci \pi \bar{k}}}{2 \sin(\pi \bar{k})}\rme^{\ci 2 \pi  x \bar{k}} \right|_{\bar{k}=k_1+k_2} 
 \partial_t\left(a_t(k_1,-\sigma) a_t(k_2,\sigma)\right)
  \nonumber \\ & \quad
 + \frac{1}{3}\sum_{\sigma\in \set{\pm 1}^3}
 \int_{\T^3}\! \rmd k\,  \Phi_3(k)
 \left.\frac{\ci \rme^{-\ci \pi \bar{k}}}{2 \sin(\pi \bar{k})}\rme^{\ci 2 \pi  x \bar{k}} \right|_{\bar{k}=\sum_{\ell=1}^3 k_\ell}   \partial_t\left(
 \prod_{\ell=1}^3 a_t(k_\ell,\sigma_\ell) \right)
  \nonumber \\ & \quad
 + \frac{1}{4}\sum_{\sigma\in \set{\pm 1}^4}
 \int_{\T^4}\! \rmd k\, \Phi_4(k)
 \left.\frac{\ci \rme^{-\ci \pi \bar{k}}}{2 \sin(\pi \bar{k})}\rme^{\ci 2 \pi  x \bar{k}} \right|_{\bar{k}=\sum_{\ell=1}^4 k_\ell}   \partial_t\left(
\prod_{\ell=1}^4 a_t(k_\ell,\sigma_\ell)\right)
\end{align}
yields a current observable associated to the energy density (\ref{eq:phononendens}) which satisfies the lattice
continuity equation (\ref{eq:latticecont}).   Let us point out that no spurious imaginary part is created by this choice: since $a_t(k,\sigma)^* = a_t(-k,-\sigma)$, $\Phi_n(k)^*=\Phi_n(-k)$, for all $n$, and $\omega(-k)=\omega(k)$, we have always 
$J_{x,x+1}(t)^*=J_{x,x+1}(t)$ and hence the current observable is real-valued.  We may also simplify
the prefactor above by using $\frac{\ci \rme^{-\ci \pi \bar{k}}}{2 \sin(\pi \bar{k})}=\frac{1}{2}\left(1+\ci\cot(\pi \bar{k})\right)$.

The above current observable also behaves as expected with respect to spatial translations of the lattice.
If we choose some $x_0\in \Z$ and define $\tilde{q}_x=q_{x+x_0}$ and $\tilde{p}_x=p_{x+x_0}$, then 
the $a$-field of the translated configuration $(\tilde{q},\tilde{p})$ satisfies
$\tilde{a}_t(k,\sigma)=\rme^{\ci 2 \pi x_0 k}a_t(k,\sigma)$. 
Thus each of the three terms in the sum defining the current 
in (\ref{eq:defencurrent}) will acquire a factor $\rme^{\ci 2 \pi  x_0 \bar{k}}$ with $\bar{k}=\sum_{\ell=1}^3 k_\ell$.
Hence, the current of the translated configuration satisfies $\tilde{J}_{x,x+1}=J_{x+x_0,x+x_0+1}$, similarly to what
was proven earlier for the energy density observables $H_x$.

\subsection{Choice of initial data}
\label{sec:initialdata}

We consider the above evolution equation 
with random initial data.  The distribution of initial data plays as important a role as the choice of the
coupling functions.  For instance, there always are degenerate initial data which do not thermalize.  Assume, for instance, that
$q(0)$ is chosen as any local minimum or maximum configuration of the potential $\mathcal{V}$ and let the particles be initially at rest, $p(0)=0$.
Since then $\nabla \mathcal{V}(q(0))=0$ this configuration is a stationary solution of the evolution equations.  However, it need not be stable, and it can well happen that even a minute perturbation of the initial data will take the asymptotic state of the system far from the original stationary state.

On the other hand, for normally conducting systems sufficiently chaotic initial data should lead to thermalization of the state.  The precise mechanism and classification of what kind of initial data should have this property, and in which precise mathematical sense, is a long-standing open problem in mathematical physics.
The most difficult part of the problem seems to be to control the beginning of the thermalization process, the \defem{microscopic thermalization}, where the state of the system approaches local stationarity.

However, many important physical properties of the system do not require full understanding of the thermalization process.  For instance, for systems with normal heat conduction, the Green--Kubo formula allows to compute the thermal conductivity of the system.  As shown in the next subsection,
the formula involves studying the time-correlations of the system \defem{starting it from a thermal equilibrium state} or, equivalently, to study the evolution of correlation functions for small perturbations of the equilibrium state.  Hence, thermal equilibrium states and their perturbations form an important class of initial data, and our assumptions should 
allow for at least these.

To arrive at the kinetic theory of phonons, we need to make some rather specific assumptions about the initial data.  The principle here is that the states are chosen to mimic states produced by typical microscopic time-averages.  
Note that this concept involves a choice of scale, the time-scale over which the time-averages are taken.  
These questions touch upon the tough question of microscopic thermalization, and we do not wish to 
speculate further about them here.  
Instead,
let us postulate a number of assumptions on the initial data based on reasonable assumptions 
about the physical characteristics of the time evolution.

We assume here that the initial data for the system has the following properties.
\begin{enumerate}
 \item {\bf Random:} We suppose that the initial data $(q_x(0),p_x(0))$ is randomly distributed.
   The initial randomness makes also the configurations $(q_x(t),p_x(t))$ at later times random variables.
 \item {\bf Chaotic:}  We suppose that initial distribution of particles in regions which are far apart are almost independent.  More precisely,
 we suppose that all correlation functions of the initial data decay fast (on the microscopic scale).  The speed is assumed to be at least so fast that the
 correlation functions are $\ell_2$-summable.
 \item {\bf Spatially homogenized:}  We assume that there is a scale $\vep^{-1}$, which is large in microscopic units, such that the correlation functions are nearly invariant under translations of lengths less than $\vep^{-1}$.
\end{enumerate}

The randomness in the initial data makes also the phonon fields 
$a_t(k,\sigma)$ random variables.  In principle, the field is not defined for $k$ such that $\omega(k)=0$, but we assume these to have been suitably regularized.  Consider for instance the case in which the interactions depend only on the differences $q_x-q_{x-1}$,
such as for the FPU-beta chains.  Given any initial data (on a finite periodic chain), we can regularize it 
before defining the phonon fields without altering the evolution equations in any way.  
We move from $(q,p)$ to $(\tilde{q},\tilde{p})$ by first setting $\bar{p} = \frac{1}{N} \sum_x p_x(0)$ and
$\bar{q} = \frac{1}{N} \sum_x q_x(0)$, and then we define $\tilde{q}_x(t) = q_x(t)-\bar{q}-t \bar{p}$ and
$\tilde{p}_x(t) = p_x(t)-\bar{p}$.  Under these assumptions, the total momentum is conserved, and thus
$\sum_x \tilde{p}_x(t)=0=\sum_x \tilde{q}_x(t)$ for all $t$.  In addition,
$(\tilde{q}_x(t),\tilde{p}_x(t))$ are a solution to the original evolution equations.
We take $a_t(k,\sigma)$ to be defined via these regularized fields which guarantees that $a_t(0,\sigma)=0$ for all $t$.

As we are interested in the evolution of the energy density (\ref{eq:phononendens}), it suffices to study the correlation functions of $a_t(k,\sigma)$.
For instance, to study the energy density averaged over the initial data, one needs the correlation functions up to order $4$.

To control the correlation functions of the above type of chaotic, spatially homogenized states, it is often
better to consider the cumulants of the fields rather than their moments.  For instance, consider
the field $p_x$ at two points $x_1, x_2$ which are far apart.  By the chaoticity assumption,
then $e_1=p^2_{x_1}$ and $e_2=p^2_{x_2}$ should be nearly independent random variables, and hence
$\mean{e_1^n e_2^m}\approx \mean{e_1^n} \mean{e_2^m}$, for $n,m\ne 0$.  Unless one of the particles is frozen to its
initial position, these moments are not zero.  However, the corresponding cumulant
is then nearly zero whenever both $n,m\ne 0$.

Hence, for two asymptotically independent regions, it is the cumulants, not moments,
which will vanish in the limit of taking the regions infinitely far from each other.
We consider here only initial states for which this decay is absolutely square summable in space,
``$\ell_2$-clustering'' for short.  This particular assumption allows for an easy identification
of the (distribution type) singularities of the Fourier-transform for any spatially homogeneous
initial data.  

To make the computations manageable, we rely here on the notations and basic results used in \cite{LM15}.
In particular, for a random field $\psi(x)$, $x\in \Z$, and any sequence $J$ of $n$ lattice points
the shorthand notations $\mean{\psi(x)^J}$, $\kappa[\psi(x_J)]$, and $\w{\psi(x)^J}$ refer to the expectation
$\mean{\prod_{i=1}^{n} \psi(x_{J_i})}$, to the cumulant $\kappa[\psi(x_{J_1}),\ldots,\psi(x_{J_n})]$,
and to the Wick polynomial $\w{\prod_{i=1}^{n} \psi(x_{J_i})}$, respectively.  In addition, we define 
$\psi(x)^J = 1$ if $J$ is an empty sequence.  Similar notations will be used for random fields on other
label sets in addition to the lattice points; for instance, for the fields $a(k,\sigma)$ and any sequence 
$J=(k_i,\sigma_i)_{i=1}^n$ we write $\kappa[a(k,\sigma)_J]=\kappa[a(k_1,\sigma_1),\ldots,a(k_n,\sigma_n)]$.

We employ here the following basic relations between the above constructions.  The proofs of these 
results can be found in many sources,
in particular, in \cite{LM15} using the above notations.  The moments-to-cumulants formula states that
for any sequence $J$
\begin{align}\label{eq:mtc}
 \mean{\psi(x)^J} = \sum_{\pi \in\mathcal{P}(J)} \prod_{A\in \pi} \kappa[\psi(x_A)]\, ,
\end{align}
where $\mathcal{P}(J)$ denotes the collection of partitions of 
the sequence $J$.\footnote{Some care is needed in the interpretation of the moments-to-cumulants formula in
order to get all combinatorial factors correctly.  It is safe to think that the elements in the 
sequence $J$ of length $n$ are labelled by the set
$I_n=\{1,2,\ldots n\}$, and the collection of partitions $\mathcal{P}(J)$ refers to the collection of standard
set partitions of $I_n$.  Therefore, if $\ell\in A\in \pi\in \mathcal{P}(J)$, we have $\ell\in I_n$, $x_\ell$
denotes the value of $x$ at the $\ell$:th position of the sequence $J$, and $x_A$ denotes the subsequence $(x_\ell)_{\ell\in A}$ of $J$.}
For a partition $\pi  \in\mathcal{P}(I)$, let us call the subsets $A\in \pi$ 
\defem{clusters} or \defem{blocks}.  We also recall that the cumulants are permutation invariant and 
multilinear, which means that 
they are linear in each of the arguments separately.  Thus cumulants of Fourier-transforms 
can be easily expressed as a linear combination of the cumulants of the original field.

Wick polynomials of random variables allow simplifying the moments-to-cumu\-lants expansions
by removing all those partitions $\pi$ from the sum which contain a cluster internal to the Wick polynomial part
of the expectation.  Explicitly, consider some $L\ge 1$ and a collection of $L+1$ index sequences
$J'$ and $J_\ell$, for $\ell=1,\ldots,L$.
Then, for the merged sequence $I=  J'+\sum_{\ell=1}^L J_\ell $ we have 
\begin{equation}\label{eq:wick_prod_multi}
\bigg\langle  \psi(x)^{J'} \prod_{\ell=1}^L \w{\psi(x)^{J_\ell} } \bigg\rangle = 
\sum_{\pi \in \mathcal{P}(I)}\prod_{A \in \pi}
\!\left(\kappa[\psi(x_A)] \cf(A \not\subset J_\ell\  \forall \ell)\right) \, .
\end{equation}
In particular, the formula implies that cumulants and Wick polynomials satisfy the following relation:
If the sequence $J$ is non-empty and has $x_1$ as its first element, then 
\begin{align}\label{eq:kappawithW}
 \kappa[\psi(x_J)]=\mean{\psi(x_1)\w{\psi_t(x)^{J\setminus x_1}}}\,,
\end{align}
where $J\setminus x_1$ denotes the sequence which is obtained when $x_1$ is removed from $J$.
By permutation invariance of the cumulants, the formula holds also if $x_1$ is replaced by any other
element in $J$.

Let us also point out that, since the cumulants are multilinear, their time-evolution can be 
written down fairly easily for the present kind of hybrid dynamics,
where all randomness is contained in the initial data.  Namely, by using a straightforward
``telescoping'' argument and (\ref{eq:kappawithW}), it is clear that for any sequence $J$
\begin{align}\label{eq:cumultimeev}
 \partial_t \kappa[\psi_t(x_J)]=\sum_{\ell\in J} \mean{\partial_t \psi_t (x_\ell)\w{\psi(x)^{J\setminus x_\ell}}}\, .
\end{align}
For the manipulation of terms such as $\partial_t \psi_t$ here, it 
is possible to move back and forth between standard and Wick polynomials by using the following identities:
\begin{align}\label{eq:inv_wick}
\psi(x)^J = \sum_{U \subset J} \w{\psi(x)^U} \mean{\psi(x)^{J \backslash U}} \, ,
\end{align}
and
\begin{align}\label{eq:wick_cum}
\w{\psi(x)^J} = \sum_{U \subset J} \psi(x)^U \sum_{\pi \in \mathcal{P}(J \backslash 
U)}(-1)^{|\pi|} \prod_{A \in \pi} \kappa[\psi(x_A)] \, ,
\end{align}
where $|\pi|$ denotes the number of clusters in the partition $\pi$.

Let us for the moment denote the inverse Fourier-transform of $\sqrt{2 \omega(k)}a(k,\sigma)$
by $\psi(x,\sigma)$.  By the definition in (\ref{eq:deffta}), we in fact then have
$\psi(x,\sigma)=\sum_y \widetilde{\omega}(x-y)q_y + \ci \sigma p_x$.  The $\ell_2$-clustering assumption
of the $(q,p)$ variables then implies that the $n$:th cumulant
$\cf(y_1=0)\kappa[\psi(x+y_1,\sigma_1),\ldots,\psi(x+y_{n},\sigma_n)]$
is square summable in $y$ for every $x$.  Thus one can safely take Fourier-transform
with respect to the variables $y$, and the result is a function in $L^2(\T^{n})$
for every fixed $x\in \Z$.  Since the second assumption implies that the function
is slowly varying in $x$, on the scale $\vep^{-1}$, we can thus conclude that
for each $\sigma\in \{\pm 1\}^n$ there is a function $F_n(x,k)$ such that 
it is slowly varying in the lattice position $x$ on the scale $\vep^{-1}$, it is $L^2$-integrable in $k\in\T^n$, and 
\begin{align}
 \kappa[\FT{\psi}(k_1,\sigma_1),\ldots,\FT{\psi}(k_n,\sigma_n)]
 = \sum_{x} \rme^{-\ci 2\pi x\cdot \sum_{\ell=1}^n k_\ell} F_n(x,k_1,k_2,\ldots,k_n;\sigma)\, .
\end{align}
In fact, there are many such functions: by using the periodic lattice $\Lambda$ as a middle step,
we find that, for instance, any convex combination of the $n$ functions, defined for $\ell_0\in \{1,\ldots,n\}$ by
\begin{align}
 & F_{n,\ell_0}(x,k_1,k_2,\ldots,k_n;\sigma)
% \nonumber \\ & \quad
 = \sum_{y\in \Lambda^n} \rme^{-\ci 2\pi \sum_{\ell=1}^n y_\ell k_\ell}
 \cf(y_{\ell_0}=0)
 \kappa[\psi(x+y_{1},\sigma_1),\ldots,\psi(x+y_{n},\sigma_n)]
 \, ,
\end{align}
will work.

In the translation invariant case, when the scale of spatial variation $\vep\to 0$,
$F_n$ is also independent of $x$ and hence can be thought of as a proper function in $L^2(\T^n)$.  Since
$\FT{\psi}(k,\sigma)=\sqrt{2 \omega(k)}a(k,\sigma)$, this yields the following
structure for cumulants of the $a$-fields for translation invariant, $\ell_2$-clustering, initial data:
\begin{align}\label{eq:ahomogparam}
 \kappa[a(k_1,\sigma_1),\ldots,a(k_n,\sigma_n)]
 = \delta\Bigl(\sum_{\ell=1}^n k_\ell\Bigr) F_n(k,\sigma)\prod_{\ell=1}^n \frac{1}{(2\omega(k_\ell))^{1/2}}
\, .
\end{align}
In case $\omega$ has zeroes, we recall our regularized definition of the field $a$ and conclude that $F_n$ 
must then also be zero at such points.  Let $W_n(k,\sigma)$ denote the values from the product 
$F_n(k,\sigma)\prod_{\ell=1}^n (2\omega(k_\ell))^{-1/2}$.  Thus, if $\omega(k_\ell)=0$ for some $\ell$,
then $W_n(k,\sigma)=0$ as well.

Therefore, although the cumulants are singular, their singularity structure is
simple, entirely encoded in the $\delta$-multiplier.
In contrast, by the moments-to-cumulants formula, the moments
satisfy for any sequence $J$ of labels
\begin{align}
\mean{a(k,\sigma)^{J}}= \sum_{\pi \in\mathcal{P}(J)} \prod_{A\in \pi}
\bigg(\delta\Bigl(\sum_{j\in A} k_j\Bigr) W_{|A|}(k_A,\sigma_A)
\biggr) \,.
\end{align}
This has ever more complicated singularity structure as the order of the moment
is increased.
(The above discussion can be made mathematically rigorous by replacing the
infinite lattice by a periodic $d$-dimensional lattice.
See \cite{NLS09} for details.)

\subsection{Green--Kubo formula}\label{sec:GKformula}

The Green--Kubo formula gives an expression for the thermal conductivity if the 
system has normal heat conduction, i.e., diffusive energy transport.
However, it can also be used to inspect anomalous transport since it always allows studying
certain relaxation characteristics of perturbations of thermal states.

Let us recall here the argument in the diffusive case, assuming that energy is the only relevant
ergodic invariant.  The basic assumption is that
after a relatively short thermalization period the macroscopic energy density 
$e(x,t)$ evolves according to the Fourier's law
\begin{align}\label{eq:nonlindiffeq}
 \partial_t e = \partial_x\left( D(e) \partial_x e \right) \, .
\end{align}
This is a nonlinear diffusion equation with a diffusion ``constant'' $D$ which depends on the value of 
energy density $e$.  The steady states are then given by microcanonical ensembles
whose only variable is the uniform energy density $\bar{e}$.

Consider next an infinite system with a steady state whose energy density is $\bar{e}$.
Perturb the state, locally near origin, so that the perturbation has finite energy.  
Then, for normally conducting
systems, one would expect that the energy density at ``infinity'' always remains equal to $\bar{e}$
and that the final state of an infinite system should be the same as the initial steady state,
since for finite systems of volume $V$ the final steady state should be of
uniform energy density $\bar{e}+O(V^{-1})$.  This implies that there should be a time 
$t_0$ after which the energy density is ever better approximated by a solution to 
the \defem{linear} diffusion equation 
\begin{align}\label{eq:lindiffeq}
 \partial_t e = D(\bar{e}) \partial^2_x e \, .
\end{align}
For a linear diffusion equation, it is possible to find out the diffusion constant
from the leading evolution of the spatial spread of the perturbation.  Namely,
since then 
\begin{align}\label{eq:gausssol}
 e(t+t_0,x) \simeq \int \rmd y \frac{1}{\sqrt{4\pi D(\bar{e})t}} \rme^{-\frac{1}{4 D(\bar{e})t} (x-y)^2}
 e(t_0,y) \, ,
\end{align}
where $\int \!\rmd x\, x^2 |e(t_0,x)-\bar{e}| <\infty$, it follows that
\begin{align}\label{eq:contexcess}
 \lim_{t\to \infty} 
 \frac{1}{t}\int \!\rmd x\, x^2 (e(t,x)-\bar{e}) = D(\bar{e}) 2 \delta E\, ,
\end{align}
where $\delta E$ denote the excess energy of the perturbation, which is conserved and hence
$\delta E=\int \!\rmd x\, (e(t,x)-\bar{e})$ for all $t\ge 0$.

Translated into the present lattice setting, one analogously arrives at the formula
\begin{align}\label{eq:discexcess}
 \lim_{t\to \infty}
 \frac{1}{t} \sum_{x\in \Z} x^2 \kappa[H_x(t),H_0(0)]
 =D(e) 2 \sum_{x\in \Z} \kappa[H_x(0),H_0(0)] \, ,
\end{align}
where $\kappa[a,b]=\mean{ab}-\mean{a}\mean{b}=\mean{(a-\mean{a})(b-\mean{b})}$ denotes the second cumulant, i.e.,
covariance, of the random variables $a,b$ defined  by solving the time-evolution using equilibrium initial data with mean energy $e$.  
Namely, if we divide the equation by $\mean{H_0}$, 
the sum on the right hand side corresponds to the excess energy from a perturbation of the initial measure $\mu_0$
to the probability measure $\frac{H_0}{\mean{H_0}} \mu_0$.  The perturbation is then localized near the origin. 
Since the left hand side is a discrete version of the left hand side of (\ref{eq:contexcess}),
the limit in (\ref{eq:discexcess}) should hold if the relaxation occurs
via diffusion of energy, as given by (\ref{eq:nonlindiffeq}).

To get the standard Green--Kubo formula, 
we next also assume equivalence of the infinite volume microcanonical and canonical ensembles, and use here
initial data distributed according to a canonical ensemble at a temperature $T$.  Then $e$ in (\ref{eq:discexcess})
denotes the corresponding (uniform) equilibrium energy density, $e=e(T)=\mean{H_0}_T$. 
Making this change of variables in (\ref{eq:nonlindiffeq}) implies the standard Fourier's law.
Explicitly, we define the temperature distribution corresponding to a given $e(x,t)$
by inverting the above function to yield $T(t,x)=T(e(t,x))$ which then satisfies
\begin{align}
 \frac{\partial e(T)}{\partial T}\partial_t T = \partial_x\left( \kappa(T) \partial_x T \right) \, ,
\end{align}
where the \defem{thermal conductivity} is given by $\kappa(T)=D(e(T))\frac{\partial e(T)}{\partial T}$.
In the finite volume $\Lambda$ canonical Gibbs state, we have $e_\Lambda(T)=
\mean{H_0}_\Lambda=Z^{-1}\int\! \rmd^\Lambda q\, \rmd^\Lambda p\,\rme^{-H/T} H_0$ where $Z$ denotes 
the partition function, $Z=\int\! \rmd^\Lambda q\, \rmd^\Lambda p\, \rme^{-H/T}$.  Therefore, 
\begin{align}
\frac{\partial e_\Lambda(T)}{\partial T}= \frac{1}{T^2} \left(\mean{H H_0}_\Lambda-\mean{H}_\Lambda\mean{H_0}_\Lambda\right)
= \frac{1}{T^2} \sum_{x\in \Lambda} \kappa_\Lambda[H_x(0),H_0(0)] \, .
\end{align}
Taking the infinite volume limit $\Lambda\to \Z$, and then using (\ref{eq:discexcess}), thus yields
\begin{align}\label{eq:GK0}
 \kappa(T) = \frac{1}{2 T^2}\lim_{t\to \infty}
 \frac{1}{t} \sum_{x\in \Z} x^2 \kappa[H_x(t),H_0(0)]\, .
\end{align}

The above argument shows that to find the numerical value of the thermal conductivity at a given temperature, it
suffices to study the asymptotic increase of the 
spatial quadratic spread of the energy density, when starting the system 
with samples taken from the corresponding canonical Gibbs state.  Explicitly, one should study the observable
\begin{align}
  S(t) = \sum_{x\in \Z} x^2 \kappa[H_x(t),H_0(0)] \,,
\end{align}
show that it is $O(t)$ for large $t$, and then solve the leading asymptotics.

For this, it is convenient to rewrite $S(t)$ in terms of current observables.  
Namely, for any $J$ satisfying (\ref{eq:latticecont}) we have for $t>0$ and all $x\in \Z$
\begin{align}
 H_x(t) & = H_x(0) + \int_0^t\!\rmd s\, \left( J_{x-1,x}(s)-J_{x,x+1}(s)\right)\, , \label{eq:Hxtint}\\
 H_x(-t) & = H_x(0) - \int_{-t}^0\!\rmd s\, \left( J_{x-1,x}(s)-J_{x,x+1}(s)\right)\, .
 \label{eq:Hxminustint}
\end{align}
Since the Gibbs state is invariant under both spatial translations and the Hamiltonian time-evolution, and 
these transform the energy observables as expected, we have here
\begin{align}
  S(t) = \sum_{x\in \Z} x^2 \kappa[H_0(0),H_{-x}(-t)] 
  = \sum_{x\in \Z} x^2 \kappa[H_{x}(-t),H_0(0)] = S(-t)\, .
\end{align}
Hence, $S(t)$ is symmetric, and
\begin{align}
 & 2S(t) = 2S(0) 
 + \sum_{x\in \Z} x^2  \int_0^t\!\rmd s\, \kappa[ J_{x-1,x}(s)-J_{x,x+1}(s),H_0(0)]
  \nonumber \\ & \qquad
 -\sum_{x\in \Z} x^2  \int_{-t}^0\!\rmd s\, \kappa[ J_{x-1,x}(s)-J_{x,x+1}(s),H_0(0)]
  \nonumber \\ & \quad
 = 2S(0)
 + \sum_{x\in \Z} (x^2-(x-1)^2)  \int_0^t\!\rmd s\, \kappa[ J_{x-1,x}(0),(H_0(-s)-H_0(s))]\, ,
\end{align}
where we have used the earlier proven simple rule for translation of the current observable.
Therefore, by (\ref{eq:Hxtint}) and (\ref{eq:Hxminustint}), we have
\begin{align}\label{eq:GreenKubo0}
 & S(t) = S(0) 
 - \frac{1}{2}\sum_{x\in \Z} (2 x -1)  \int_0^t\!\rmd s\, \int_{-s}^s\!\rmd r\, 
 \kappa[J_{x-1,x}(0), J_{-1,0}(r) -J_{0,1}(r)]
  \nonumber \\ & \quad 
 = S(0) 
 + \int_0^t\!\rmd s\, \int_{-s}^s\!\rmd r\, 
   \sum_{x\in \Z} \kappa[ J_{x-1,x}(0),J_{0,1}(r)]\, .
\end{align}

We define the \defem{current-current correlation function},
also called the \defem{Green--Kubo correlator}, by the formula
$C(t;T) = \sum_{x\in \Z} \kappa[ J_{x-1,x}(0),J_{0,1}(t)] $.
From the above properties, we obtain $C(-t) = \sum_{x\in \Z} \kappa[ J_{x-1,x}(t),J_{0,1}(0)] =
\sum_{x\in \Z} \kappa[ J_{0,1}(t),J_{1-x,2-x}(0)] = C(t)$, and thus
we find from (\ref{eq:GreenKubo0})
\begin{align}\label{eq:finalStform}
 & S(t) = S(0) 
 + 2 \int_0^t\!\rmd s\, \int_{0}^s\!\rmd r\, C(r)
 = S(0) 
 + 2 t \int_0^t\!\rmd r\, \left(1-\frac{r}{t}\right)C(r) \, .
\end{align}
If $\int_0^\infty \!\rmd r\, |C(r;T)|<\infty$, we can rely on dominated convergence in (\ref{eq:finalStform})
and conclude that then, by (\ref{eq:GK0}),
\begin{align}\label{eq:GK1}
 \kappa(T) = \frac{1}{T^2} \int_0^\infty\!\rmd r\, C(r;T)\, .
\end{align}
This is the standard Green--Kubo formula, and thus, in great generality, the integrability of the correlation function $C(t;T)$
is sufficient to imply normal thermal conduction at the temperature $T$.  

However, we can also conclude that, if
$\int_0^t\!\rmd r\, \left(1-\frac{r}{t}\right)C(r;T)$ is \defem{not} bounded in $t$, 
then the energy relaxation cannot be diffusive at temperature $T$.
Hence, the Green--Kubo correlation function $C(t;T)$ can also be used to prove 
\defem{superdiffusion} of energy and to check at which time scale the energy spread occurs: if 
$\int_0^t\!\rmd r\, \left(1-\frac{r}{t}\right)C(r;T)=O(t^{p})$, with $p>0$, then the spatial spread at time $t$
is $O(t^{(1+p)/2})$.   However, the spread from the observable $S(t)$ can be somewhat misleading for systems which have 
polynomial decay or multiscale behaviour.  In these cases, $S(t)$ could be entirely dominated by tail behaviour, 
or just by one of the many scales, 
instead of describing the actual speed of spreading of local perturbations.
We will find such an example from the kinetic theory of FPU-$\beta$ chains, discussed in Sec.~\ref{sec:CtFPU}.

Let us conclude the section by computing an expression for the current correlation $C(t)$ for the
particle chains.  Some care needs to be taken with the sum over $x$ and the principal value integral 
in the definition of the current observable.  To control their behaviour,
let us insert $\int_{\T} \!\rmd \bar{k}\, \delta(\bar{k}-\sum_\ell k_\ell)$ into the integrals defining $J$ in (\ref{eq:defencurrent}).
Then for each $t$ there is some function $f_t$ on the torus, such that 
\begin{align}
 \kappa[ J_{x-1,x}(0),J_{0,1}(t)] = \int_{\T} \!\rmd \bar{k}\, \rme^{\ci 2\pi x \bar{k}}
 \frac{1}{2}\left(1+\ci\cot(\pi \bar{k})\right) f_t(\bar{k})\,, 
\end{align}
as a principal value integral around zero.  Hence,
\begin{align}
& \sum_{x\in\Z} \kappa[ J_{x-1,x}(0),J_{0,1}(t)] = \frac{1}{2} f_t(0)
 + \ci\sum_{x\in\Z} \int_{\T} \!\rmd \bar{k}\, \rme^{\ci 2\pi x \bar{k}}
 \frac{f_t(\bar{k})-f_t(-\bar{k})}{4}\cot(\pi \bar{k})
  \nonumber \\ & \quad 
 = \frac{1}{2} f_t(0)
 + \frac{\ci}{2\pi} \partial_{\bar{k}}f_t(0)\,.
\end{align}
In fact, here $f_t(0)=0$ due to energy conservation, as discussed above.
Also, for any function $F$ we have $\left.\partial_{\bar{k}} \int_{\T^n}\! \rmd k \, \delta(\bar{k}-\sum_\ell k_\ell)
F(k)\right|_{\bar{k}=0}=\int_{\T^n}\! \rmd k \, \delta(\sum_\ell k_\ell)
\partial_{k_1} F(k)$.

Therefore, for the particle chains the Green--Kubo correlation function is given by
\begin{align}
& \sum_{x\in\Z} \kappa[ J_{x-1,x}(0),J_{0,1}(t)] = 
\sum_{n',n=2}^4 \frac{1}{n m} \sum_{\sigma',\sigma} w_n(\sigma) w_{n'}(\sigma') \int_{\T^n}\! \rmd k \, 
\delta\!\left(\sum_{\ell=1}^n k_\ell\right) \int_{\T^{n'}}\! \rmd k' \, 
  \nonumber \\ & \qquad  \times
\frac{\ci}{4\pi} \Phi_{n'}(k') \Bigl(-1+\ci\cot(\pi \sum_{\ell=1}^{n'} k'_\ell)\Bigr)
  \nonumber \\ & \qquad  \times
\kappa\!\Bigl[\partial_{k_1}\! \Bigl[\Phi_n(k)\partial_s\Bigl(\prod_{\ell=1}^n a_s(k_\ell,\sigma_\ell)\Bigr)_{s=0}\Bigr],
 \partial_t\Bigl(\prod_{\ell=1}^{n'} a_t(k'_\ell,\sigma'_\ell)\Bigr)\Bigr]\, ,
\end{align}
where the weight $w_n(\sigma)=1$, for $\sigma \in \{\pm 1\}^n$, unless $n=2$ and $\sigma_1=\sigma_2$, in which case it is zero.  
By translation invariance, the final cumulant is proportional to $\delta(\sum_{\ell=1}^n k_\ell+\sum_{\ell=1}^{n'} k'_\ell)$,
and thus the integral over $k'$ has to be treated similarly to what has been done above for the integral over $k$.
Following the same steps as above, we thus arrive at the simplified formula
\begin{align}
& \sum_{x\in\Z} \kappa[ J_{x-1,x}(0),J_{0,1}(t)] =
\sum_{n',n=2}^4 \sum_{\sigma' \in \{\pm 1\}^{n'}} \sum_{\sigma \in \{\pm 1\}^n}w_n(\sigma) w_{n'}(\sigma') 
  \nonumber \\ & \qquad  \times
\int_{\T^n}\! \rmd k \, \delta\!\left(\sum_{\ell=1}^n k_\ell\right) \int_{\T^{n'}}\! \rmd k' \, 
\kappa\!\Bigl[\mathcal{J}_n(k,\sigma;0),\mathcal{J}_{n'}(k',\sigma';t)\Bigr] \, ,
\end{align}
where for $k\in \T^n$, $\sigma \in\{\pm 1\}^n$, we have defined
\begin{align}
 \mathcal{J}_n(k,\sigma;t) = \frac{1}{n}\frac{\ci}{2\pi}\partial_{k_1}\! \Bigl[\Phi_n(k)\partial_t\Bigl(\prod_{\ell=1}^n a_t(k_\ell,\sigma_\ell)\Bigr)\Bigr]
 \, .
\end{align}

Assuming that the harmonic terms dominate over 
the anharmonic ones, we can obtain the leading harmonic part of the above expression
by considering only the term $n'=2=n$ above and approximating $\partial_t a_t(k,\sigma)\approx -\ci \sigma\omega(k)a_t(k,\sigma)$.
This yields an approximation
\begin{align}
 \mathcal{J}_2(k,\sigma;t) \approx \frac{1}{4\pi}\partial_{k_1}\! \Bigl[\Phi_n(k) (\sigma_1 \omega(k_1)+\sigma_2\omega(k_2))
 a_t(k_1,\sigma_1) a_t(k_2,\sigma_2)\Bigr]
 \, .
\end{align}
Since these terms appear in the integrand only when $\sigma_2=-\sigma_1$ and $k_2=-k_1$, for which 
$\sigma_1 \omega(k_1)+\sigma_2\omega(k_2)=0$, only the term where the factor $\sigma_1 \omega(k_1)+\sigma_2\omega(k_2)$
is differentiated yields a non-zero contribution
to the integral.  Since then also $\Phi_2(k)=\omega(k_1)$, it is possible to substitute the above harmonic current term by 
\begin{align}
\omega(k_1) \frac{\sigma_1}{4\pi}\omega'(k_1)
 a_t(k_1,\sigma_1) a_t(k_2,\sigma_2)
\end{align}
in the integrand.
Thus the harmonic part of the Green--Kubo correlation function evaluates to 
\begin{align}\label{eq:harmcurrent}
& \sum_{\sigma',\sigma \in \{\pm 1\} }  \frac{\sigma'\sigma}{4}
\int_{\T}\! \rmd k \int_{\T^2}\! \rmd k' \, 
\frac{\omega'(k)}{2\pi} \omega(k)  \frac{\omega'(k'_1)}{2\pi} \omega(k'_1)  
 % \nonumber \\ & \quad  \times
\kappa\!\Bigl[ a_0(k,\sigma) a_0(-k,-\sigma), a_t(k'_1,\sigma') a_t(k'_2,-\sigma')\Bigr]\, .
\end{align}

As $\omega'(-k)=-\omega'(k)$, we can simplify the expression (\ref{eq:harmcurrent}) further to 
\begin{align}
& 
\int_{\T}\! \rmd k \int_{\T^2}\! \rmd k' \, 
  %\nonumber \\ & \quad  \times
\frac{\omega'(k)}{2\pi} \omega(k)  \frac{\omega'(k'_2)}{2\pi} \omega(k'_2)  
\kappa\!\Bigl[ a_0(-k,-1) a_0(k,1), a_t(k'_1,-1) a_t(k'_2,1)\Bigr]\, .
\end{align}
Here $\frac{\omega'(k)}{2\pi} \omega(k) = \frac{1}{4\pi}\FT{\alpha}'(k)$ which is equal to 
$\ombasic^2 \delta \sin (2\pi k)$
for chains with stable 
nearest neighbour interactions, using the parametrization given in Section \ref{sec:freephonons}.
Thus the harmonic interaction part of the Green--Kubo correlation function for
nearest neighbour chains is equal to
\begin{align}\label{eq:Ctharm2}
& \ombasic^4 \delta^2
\kappa\!\left[ 
\int_{\T}\! \rmd k \sin (2\pi k)
a_0(-k,-1) a_0(k,1), \int_{\T^2}\! \rmd k' \,\sin (2\pi k_2') a_t(k'_1,-1) a_t(k'_2,1)\right]\, .
\end{align}

This result coincides with the formula used in Section 3 in \cite{ALS06}.  
As explained there, the above time-correlator can also be computed by solving
the evolution of the Wigner function starting from a suitably perturbed initial Gibbs state.
Namely, consider the Gibbs states with perturbed Hamiltonians
$H^\vep(q,p)= H(q,p)-T \vep \mathcal{J}\!(q,p)$ where 
\begin{align}
 \mathcal{J}\!(q,p) = \int_{\T}\! \rmd k \sin (2\pi k) a_0(-k,-1) a_0(k,1)\,
\end{align}
and denote the expectation over the canonical Gibbs state at temperature $T$
and the Hamiltonian $H^\vep$ by $\mean{\cdot}^{(\vep)}$.
For instance by using approximations by finite periodic lattices,
it follows that
\begin{align}\label{eq:vepev}
& \left.\partial_\vep \left\langle\int_{\T^2}\! \rmd k' \,\sin (2\pi k_2') a_t(k'_1,-1) a_t(k'_2,1)
\right\rangle^{(\vep)}\right|_{\vep=0}
\nonumber \\ & \quad  
=
\kappa\!\left[ 
\int_{\T}\! \rmd k \sin (2\pi k)
a_0(-k,-1) a_0(k,1), \int_{\T^2}\! \rmd k' \,\sin (2\pi k_2') a_t(k'_1,-1) a_t(k'_2,1)\right]\, .
\end{align}
Note that the \defem{time-evolution} in both expectations is the same, determined by the 
original Hamiltonian $H$.  Hence, computing 
the left hand side involves solving the original problem with 
non-stationary initial data.  The perturbation in the initial data
can be captured by changing $\beta \omega(k)$ to $\beta \omega(k)-\vep \sigma\sin (2\pi k)$
in the harmonic part multiplying $a_0(k,-\sigma) a_0(k,\sigma)$.  In
particular, also the perturbed initial state is translation invariant in space.

\section{Phonon Boltzmann equation of spatially homogeneous anharmonic chains}

\subsection{Identifying the collision operator}
\label{sec:deriveC}

The goal of the section is to derive the form of the phonon Boltzmann collision operators for the particle chains.
We  do this only for spatially homogeneous initial data which suffices to cover the evolution
of the Green--Kubo correlation function, as explained in Sec.~\ref{sec:GKformula}.
Allowing for inhomogeneous initial data would be an important extension, but the inhomogeneity introduces
new multiscale effects which we wish to avoid in the present contribution.  See the discussion
in Section 6 of \cite{LM15} for an example of the problems associated with spatially inhomogeneous
initial data for a nonlinear evolution equation similar to the FPU chains.

Let us first recall the parametrization of the spatially homogeneous correlation functions 
$W_{n}(k,\sigma;t)$ satisfying (\ref{eq:ahomogparam}), 
\begin{align}%\label{eq:ahomogparam}
 \kappa[a_t(k_1,\sigma_1),\ldots,a_t(k_n,\sigma_n)]
 = \delta\Bigl(\sum_{\ell=1}^n k_\ell\Bigr) W_n(k,\sigma;t)
\, .
\end{align}
As discussed earlier, this requirement does not fix the functions $W_n$ completely, and here we 
resolve the ambiguity by requiring that they are constant in the first variable, $k_1$.  This is equivalent
to using the definition 
\begin{align}%\label{eq:ahomogparam}
 W_n(k,\sigma;t) = \int_{\T}\!\rmd k'_1\, \kappa[a_t(k'_1,\sigma_1),a_t(k_2,\sigma_2),\ldots,a_t(k_n,\sigma_n)]
\, .
\end{align}
We can then also obtain the Wigner function $W(k;t)$ of the state, as used in earlier works,
by setting $W(k;t)=W_2((k',k),(-1,1);t)$ for all $k',k\in \T$.

The time-evolution of the correlation functions $W_n$ can now be solved from (\ref{eq:cumultimeev}).
Recalling the earlier notations involving a sequence $J$ of labels, we have
\begin{align}
 \partial_t  \kappa[a_t(k,\sigma)_J] = \sum_{\ell \in J} \mean{\partial_t a_t(k_\ell,\sigma_\ell) \w{a_t(k,\sigma)^{J\setminus \ell}}}\, .
\end{align}
Since the derivative of the phonon field satisfies (\ref{eq:atevolanh}), we thus obtain 
\begin{align}%\label{eq:atevolanh}
& \partial_t  \kappa[a_t(k,\sigma)_J] = -\ci \Omega_n(k,\sigma)_J\kappa[a_t(k,\sigma)_J] 
 \nonumber \\ & \qquad
 -\ci \sum_{\ell\in J} \sigma_\ell \sum_{\sigma'\in \set{\pm 1}^2}
 \int_{\T^2}\! \rmd k'\, \delta(k_\ell-k'_1-k'_2) \Phi_3(k_\ell,k') 
 \nonumber \\ & \qquad\quad \times
 \mean{ a_t(k'_1,\sigma'_1) a_t(k'_2,\sigma'_2) \w{a_t(k,\sigma)^{J\setminus \ell}}}
 \nonumber \\ & \qquad
 -\ci \sum_{\ell\in J} \sigma_\ell \sum_{\sigma'\in \set{\pm 1}^3}
 \int_{\T^3}\! \rmd k'\, \delta(k_\ell-\sum_{i=1}^3 k'_i) \Phi_4(k_\ell,k')
 \nonumber \\ & \qquad\quad \times
 \mean{ a_t(k'_1,\sigma'_1) a_t(k'_2,\sigma'_2) a_t(k'_3,\sigma'_3)\w{a_t(k,\sigma)^{J\setminus \ell}}}
 \, ,
\end{align}
where $n=|J|$ and we have used (\ref{eq:kappawithW}) to simplify the linear, free evolution, term into a
multiplication by the function 
\begin{align}
 \Omega_n(k,\sigma) = \sum_{\ell=1}^n \sigma_\ell \omega(k_\ell)\, .
\end{align}
The remaining expectation values can be represented in terms of cumulants using (\ref{eq:wick_prod_multi}).
To derive the Boltzmann equation, we will need only the first four cumulants but let us postpone the detailed
expansion for later.

We next exponentiate the free evolution part by integrating
the identity
\begin{align}
 & \partial_s\left(\rme^{-\ci \Omega_n(k,\sigma)_J (t-s)} \kappa[a_s(k,\sigma)_J] \right) =
  \nonumber \\ & \quad
 = \rme^{-\ci \Omega_n(k,\sigma)_J (t-s)} \left(\ci \Omega_n(k,\sigma)_J \kappa[a_s(k,\sigma)_J]+ \partial_s\kappa[a_s(k,\sigma)_J] \right)\, .
\end{align}
This results in the following Duhamel perturbation formula
\begin{align}\label{eq:knpertform}
 &  \kappa[a_t(k,\sigma)_J]
 = \rme^{-\ci \Omega_n(k,\sigma)_J t} \kappa[a_0(k,\sigma)_J]
  \nonumber \\ & \qquad
 -\ci \sum_{\ell\in J} \sigma_\ell \int_0^t\!\rmd s\, \rme^{-\ci \Omega_n(k,\sigma)_J (t-s)}
   \sum_{\sigma'\in \set{\pm 1}^2}
 \int_{\T^2}\! \rmd k'\, \delta(k_\ell-k'_1-k'_2) \Phi_3(k_\ell,k') 
 \nonumber \\ & \qquad\quad \times
 \mean{ a_s(k'_1,\sigma'_1) a_s(k'_2,\sigma'_2) \w{a_s(k,\sigma)^{J\setminus \ell}}}
 \nonumber \\ & \qquad
  -\ci \sum_{\ell\in J} \sigma_\ell \int_0^t\!\rmd s\, \rme^{-\ci \Omega_n(k,\sigma)_J (t-s)} \sum_{\sigma'\in \set{\pm 1}^3}
 \int_{\T^3}\! \rmd k'\, \delta(k_\ell-\sum_{i=1}^3 k'_i) \Phi_4(k_\ell,k')
 \nonumber \\ & \qquad\quad \times
 \mean{ a_s(k'_1,\sigma'_1) a_s(k'_2,\sigma'_2) a_s(k'_3,\sigma'_3)\w{a_s(k,\sigma)^{J\setminus \ell}}}
\, .
\end{align}

The case with $n=2$ and $J=((k_1,-1),(k_2,1))$ determines the Wigner function $W(k;t)$ after integration over $k_1$.
However, in this case $\Omega_n(k,\sigma)_J = \omega(k_2)-\omega(k_1)$ which is identically zero whenever $k_1+k_2=0$.
By translation invariance, all three terms on the right hand side of (\ref{eq:knpertform}) are now proportional 
to $\delta(k_1+k_2)$, the last two since the remaining expectations are proportional to $\delta(k_{\ell'}+\sum_{i} k'_i)$
with $(\ell')=J\setminus \ell$.
Therefore, all the phase factors involving $\Omega_n$ are equal to one in this case.

Let us next assume that the fields have been centred, i.e., that $\mean{a_t(k,\sigma)}=0$.  In general,
for a spatially homogeneous case we could have $\mean{a_t(k,\sigma)}=\delta(k) c_t(\sigma)$.  However,
for instance in the FPU model we have normalized the fields so that $a_t(0,\sigma)=0$ for all $t$, which implies $c_t(\sigma)=0$.
Under this simplifying assumption, we obtain using (\ref{eq:wick_prod_multi}) and the obvious shorthand 
notations for the fields $a'_s$ that
\begin{align}
  \mean{ a'_1 a'_2 \w{a} } & = \kappa[a'_1, a'_2, a]\, ,
  \\ 
  \mean{ a'_1 a'_2 a'_3 \w{a} }&  = \kappa[a'_1, a'_2, a'_3, a]
% \nonumber \\ & \qquad
  + \kappa[a'_1, a'_2]\kappa[a'_3, a] + \kappa[a'_1, a'_3]\kappa[a'_2, a]
  + \kappa[a'_2, a'_3]\kappa[a'_1, a] \, ,
\end{align}
where $a=a_s(k_{\ell'},\sigma_{\ell'})$. Therefore,
integration of (\ref{eq:knpertform}) over $k_1$ and writing $k=k_2$ yields
\begin{align}\label{eq:WFPUpert}
 &
  W(k;t) = W(k;0)
  \nonumber \\ & \quad
 -\ci \int_0^t\!\rmd s\, 
   \sum_{\sigma'\in \set{\pm 1}^3} \sigma'_3
 \int_{\T^3}\! \rmd k'\, \delta(k'_3-\sigma'_3 k) \delta(\sum_{i=1}^3 k'_i) \Phi_3(k') 
% \nonumber \\ & \qquad \times
 W_3(k',\sigma';s)
% \mean{ a_s(k'_1,\sigma'_1) a_s(k'_2,\sigma'_2) \w{a_s(k_{3-\ell},-\sigma_{\ell})}}
  \nonumber \\ & \quad
 -\ci \int_0^t\!\rmd s\, 
   \sum_{\sigma'\in \set{\pm 1}^4} \sigma'_4
 \int_{\T^4}\! \rmd k'\, \delta(k'_4-\sigma'_4 k) \delta(\sum_{i=1}^4 k'_i) \Phi_4(k') 
% \nonumber \\ & \qquad \times
 W_4(k',\sigma';s)
 \nonumber \\ & \quad
 -3 \ci \int_0^t\!\rmd s\, 
   \sum_{\sigma'\in \set{\pm 1}^4} \sigma'_4
 \int_{\T^4}\! \rmd k'\, \delta(k'_4-\sigma'_4 k) \delta(\sum_{i=1}^4 k'_i)  \delta(k'_1+k'_2)
\Phi_4(k') 
 \nonumber \\ & \qquad \times
 W_2((k'_1,k'_2),(\sigma'_1,\sigma'_2);s) 
 W_2((k'_3,k'_4),(\sigma'_3,\sigma'_4);s)
\, ,
 \end{align}
where we have enumerated the sum over $\ell\in J$ by using $\sigma'_{n}=-\sigma_\ell$,
when clearly $k_\ell=-\sigma'_3 k$ and $k_{\ell'}=\sigma'_3 k$.
We have also used here the 
permutation invariance of $\Phi_n$ and the obvious antisymmetries of $\Phi_n$, $n=3,4$,
under a reversal of the sign of the first component.  

The last term in (\ref{eq:WFPUpert}) can be simplified further using symmetries of $\Phi_4$ to 
\begin{align}
& -3 \ci \int_0^t\!\rmd s\, 
   \sum_{\sigma\in \set{\pm 1}^2} \sigma_2
    W_2((-\sigma_2 k,\sigma_2 k),(\sigma_1,\sigma_2);s)
 \nonumber \\ & \qquad \times
 \int_{\T}\! \rmd k'\, \Phi_4(-k',k',-k,k) 
    \sum_{\sigma'\in \set{\pm 1}^2} 
W_2((-k',k'),(\sigma'_1,\sigma'_2);s) 
\, .
\end{align}
Inserting the definition of the FPU interaction amplitude $\Phi_4$ given in (\ref{eq:defPhiFPU4})
then yields
\begin{align}
 & \ci 6 \lambda_4 \ombasic^{-2} \int_0^t\!\rmd s\,  |\sin(\pi k)|
    \sum_{\sigma\in \set{\pm 1}^2} \sigma_2
    W_2((-\sigma_2 k,\sigma_2 k),(\sigma_1,\sigma_2);s)
 \nonumber \\ & \quad \times
 \int_{\T}\! \rmd k'\, |\sin(\pi k')|
    \sum_{\sigma'\in \set{\pm 1}^2} 
W_2((-k',k'),(\sigma'_1,\sigma'_2);s) \, .
\end{align}
Since now $\ombasic |\sin(\pi k)|=\omega(k)/\sqrt{2}$ and 
% $ \sum_{\sigma\in \set{\pm 1}^2} \sigma_2
%     W_2((-\sigma_2 k,\sigma_2 k),(\sigma_1,\sigma_2);s)
%  = W_2((-k,k),(1,1);s)-W_2((-k,k),(-1,-1);s)$,
\begin{align*}
 &    \sum_{\sigma\in \set{\pm 1}^2} \sigma_2
    W_2((-\sigma_2 k,\sigma_2 k),(\sigma_1,\sigma_2);s)
% \nonumber \\ & \quad 
 = W_2((-k,k),(1,1);s)-W_2((-k,k),(-1,-1);s)
\, ,
\end{align*}
for the FPU chains we may write this term as 
\begin{align}\label{eq:extraWW}
 & \ci 3 \lambda_4 \ombasic^{-4}\int_0^t\!\rmd s\, 
 \omega(k) \sum_{\sigma\in \set{\pm 1}} \sigma W_2((-k,k),(\sigma,\sigma);s)
 \nonumber \\ & \quad \times
 \int_{\T}\! \rmd k'\, \omega(k')
    \sum_{\sigma'\in \set{\pm 1}^2} 
W_2((-k',k'),(\sigma'_1,\sigma'_2);s) \, .
\end{align}

It is customary in kinetic theory to ignore the term (\ref{eq:extraWW}) in the Boltzmann equation.
At the first glance, it is not obvious why this should be so: the term does not involve
any non-Gaussian factors and it has an apparent magnitude $O(\lambda_4 t)$.
Although no full mathematical study has been made about the effect of the term on the kinetic time scales
$O(\lambda_4^{-2})$, the following argument suggests that, indeed, it behaves better than suggested
by its apparent magnitude which is $O(\lambda_4^{-1})$ on the kinetic time scale.  
In fact, the argument indicates that the term is $O(\lambda_4)$
uniformly in $t$ and $k$, and hence could be safely neglected when $\lambda_4\ll 1$.

We study the two $W_2$-dependent factors in (\ref{eq:extraWW}) separately.  
The second factor is
\begin{align}
& \int_{\T}\! \rmd k\, \omega(k)
    \sum_{\sigma\in \set{\pm 1}^2}
W_2((-k,k),(\sigma_1,\sigma_2);s)
 \nonumber \\ & \quad
 = \int_{\T^2}\! \rmd^2 k\, \sqrt{\omega(k_1)\omega(k_2)}
    \sum_{\sigma\in \set{\pm 1}^2}
    \kappa[a_s(k_1,\sigma_1),a_s(k_2,\sigma_2)]
 \nonumber \\ & \quad
 = 2 \int_{\T^2}\! \rmd^2 k\, \omega(k_1)\omega(k_2)
    \mean{\FT{q}(s,k_1)\FT{q}(s,k_2)} \, ,
\end{align}
where we have applied the inversion formula given in (\ref{eq:invphonon}) and the assumption that $\mean{a_s(k,\sigma)}=0$.
This is equal to $2 \mean{(\widetilde{\omega}*q(s))(0)^2}$ which for a finite periodic lattice would correspond to
$2|\Lambda|^{-1} \sum_x \mean{(\widetilde{\omega}*q(s))(x)^2} =  2|\Lambda|^{-1} \mean{\sum_{x,y} q_x(s) \alpha(x-y)q_y(s)}$,
by translation invariance.  Therefore, the second factor is proportional to 
the average harmonic potential energy, and we would expect it to be slowly varying and to equilibrate to the corresponding 
equilibrium value as $s\to \infty$.  In particular, it can be bounded by the total energy density which is a constant, so the value
of this term is $O(1)$ uniformly in $s$.

The first factor in (\ref{eq:extraWW}) depends only on the ``off-diagonal'' elements of the second order correlation matrix.
By (\ref{eq:knpertform}), they have rapidly oscillating phase factors $\rme^{-\ci s \Omega}$ where
$\Omega=2\sigma \omega(k)$.  If we assume that this oscillatory behaviour continues for all $s$, it seems reasonable to expect that the value of
(\ref{eq:extraWW}) can be well approximated by using this oscillatory factor and assuming that the remainder is so slowly varying that it can be replaced 
by a constant.  This yields a factor which is $O(\lambda_4)$ times the integral $\Omega\int_0^t\!\rmd s\, \rme^{-\ci s \Omega}=\ci (1-\rme^{-\ci t \Omega})$ which 
is bounded by $2$ for all $t$ and $k$.
Of course, the replacement of a slowly varying term by a constant is not totally accurate here, but it seems reasonable to assume that
under the present conditions the resulting corrections would also be $O(\lambda_4)$, at least up to the kinetic time scales.

The rest of the argument leading to the phonon Boltzmann collision operator is fairly standard.  The clusterings relevant to the third cumulant
are
\begin{align}
 & \mean{ a'_1 a'_2 \w{a_1 a_2} } = \kappa[a'_1, a'_2, a_1, a_2]
 %\nonumber \\ & \qquad
  + \kappa[a'_1, a_1]\kappa[a'_2, a_2] 
  + \kappa[a'_1, a_2]\kappa[a'_2, a_1]  \, ,
\end{align}
and $ \mean{ a'_1 a'_2 a'_3 \w{a_1 a_2} } $ whose expansion has only terms containing higher order cumulant factors.
We continue to assume that the oscillations of the higher order cumulants are sufficient to suppress any term which involves
integrals of $W_n$, $n>2$, over any of its arguments.
The clusterings relevant to the fourth cumulant are
$\mean{ a'_1 a'_2 \w{a_1 a_2 a_3} }$, which produces only terms containing higher order cumulant factors, and 
\begin{align}
 & 
  \mean{ a'_1 a'_2 a'_3 \w{a_1 a_2 a_3} } =
% \nonumber \\ & \qquad
    \sum_{\pi\in \text{Perm}(3)} \kappa[a'_{\pi(1)}, a_1]\kappa[a'_{\pi(2)}, a_2] \kappa[a'_{\pi(3)}, a_3]
   + \text{(higher order terms)}
  \, ,
\end{align}
where the sum goes through all permutations $\pi$ of three elements.

In summary, we then obtain the following approximations from the third and fourth cumulants,
\begin{align}%\label{eq:knpertform}
 &  W_3(k,\sigma;t) 
 \approx \rme^{-\ci \Omega_3(k,\sigma) t}  W_3(k,\sigma;0) 
  \nonumber \\ & \quad
 -2 \ci \sum_{\ell=1}^3 \sigma_\ell \int_0^t\!\rmd s\, \rme^{-\ci \Omega_3(k,\sigma) (t-s)}
    \Phi_3(k) 
% \nonumber \\ & \qquad\quad \times
  \prod_{i\ne \ell}  \sum_{\sigma'\in \set{\pm 1}} W_2((-k_i,k_i),(\sigma',\sigma_i);s)
% \sum_{\sigma'\in \set{\pm 1}^2} \left.  W_2((-k_i,k_i),(\sigma'_1,\sigma_i)) W_2((-k_j,k_j),(\sigma'_2,\sigma_j)) \right|_{(k_i,k_j,k_\ell)=k}
\, ,\\ 
 &  W_4(k,\sigma;t) 
 \approx \rme^{-\ci \Omega_4(k,\sigma) t}  W_4(k,\sigma;0) 
  \nonumber \\ & \quad
  +6 \ci \sum_{\ell=1}^4 \sigma_\ell \int_0^t\!\rmd s\, \rme^{-\ci \Omega_4(k,\sigma) (t-s)}
    \Phi_4(k) 
  \prod_{i\ne \ell}  \sum_{\sigma'\in \set{\pm 1}} W_2((-k_i,k_i),(\sigma',\sigma_i);s)
%  \nonumber \\ & \qquad\quad \times
%  \sum_{\sigma'\in \set{\pm 1}^2} \prod_{i\ne \ell}  W_2((-k_i,k_i),(\sigma'_i,\sigma_i))
\, ,
\end{align}
where we have simplified the result by applying the invariance properties of the amplitude factors $\Phi_3$ and $\Phi_4$.
We insert these approximations in (\ref{eq:WFPUpert}).  The terms depending on the cumulants at time zero are again assumed to be
negligible since they involve highly oscillatory integrands.

Since $\sum_{\sigma'\in \set{\pm 1}} W_2((-k_i,k_i),(\sigma',\sigma_i);s)$ is equal to $W(\sigma_i k_i,s)$ plus a rapidly oscillatory $(\sigma_i,\sigma_i)$-term,
this results in the approximation
\begin{align}\label{eq:pBE1}
 &
  W(k;t) \approx W(k;0)
  \nonumber \\ & \quad
 + 2 \int_0^t\!\rmd s\, \sum_{\sigma\in \set{\pm 1}^3} \sigma_3
 \int_{\T^3}\! \rmd k'\, \delta(k'_3-\sigma_3 k) \delta(\sum_{i=1}^3 k'_i) (-\Phi_3(k')^2)
\int_s^t\!\rmd t'\, \rme^{-\ci \Omega_3(k',\sigma) (t'-s)}
 \nonumber \\ & \qquad \times
\sum_{\ell=1}^3 \sigma_\ell \prod_{i\ne \ell} W(\sigma_i k'_i,s)
% \nonumber \\ & \qquad\quad \times
  \nonumber \\ & \quad
 +6 \int_0^t\!\rmd s\, 
   \sum_{\sigma\in \set{\pm 1}^4} \sigma_4
 \int_{\T^4}\! \rmd k'\, \delta(k'_4-\sigma_4 k) \delta(\sum_{i=1}^4 k'_i) \Phi_4(k')^2 
\int_s^t\!\rmd t'\, \rme^{-\ci \Omega_4(k',\sigma) (t'-s)}
 \nonumber \\ & \qquad \times
\sum_{\ell=1}^4 \sigma_\ell \prod_{i\ne \ell} W(\sigma_i k'_i,s)
\, ,
\end{align}
where we have used Fubini's theorem to exchange the order of the time-integrals.
If we swap the sign of both $\sigma$ and $k'$, clearly  also $\Omega_n(k',\sigma)$, $n=3,4$,
change their sign.  Therefore, in the above formula, we may replace both of the terms
$\int_s^t\!\rmd t'\, \rme^{-\ci(t'-s) \Omega}$ by 
$\frac{1}{2}\int_{-(t-s)}^{t-s}\!\rmd r\, \rme^{-\ci r \Omega}\approx \pi \delta(\Omega)$, for large $t$. 

Inserting the explicit expressions for $\Phi_n$ thus results in the formula
\begin{align}\label{eq:pBE2}
 &
  W(k;t) \approx W(k;0)
  \nonumber \\ & \quad
 + 2^{\frac{5}{2}} \pi \lambda_3^2 \ombasic^{-3} \int_0^t\!\rmd s\, \sum_{\sigma\in \set{\pm 1}^3}
 \int_{\T^3}\! \rmd k'\, \delta(k'_3- k) 
 \delta(\sum_{i=1}^3 \sigma'_i k'_i) \delta(\sum_{i=1}^3 \sigma_i \omega(k'_i))
 \nonumber \\ & \qquad \times
 \prod_{\ell=1}^3 |\sin(\pi k'_\ell)|
\sum_{\ell=1}^3  \sigma_3 \sigma_\ell \prod_{i\ne \ell} W(k'_i,s)
% \nonumber \\ & \qquad\quad \times
  \nonumber \\ & \quad
 +24 \pi \lambda_4^2\ombasic^{-4}  \int_0^t\!\rmd s\, 
   \sum_{\sigma\in \set{\pm 1}^4} 
 \int_{\T^4}\! \rmd k'\, \delta(k'_4-k) 
 \delta(\sum_{i=1}^4 \sigma'_i k'_i) \delta(\sum_{i=1}^4 \sigma_i \omega(k'_i))
 \nonumber \\ & \qquad \times
 \prod_{\ell=1}^4 |\sin(\pi k'_\ell)|
\sum_{\ell=1}^4 \sigma_4 \sigma_\ell \prod_{i\ne \ell} W(k'_i,s)
\, .
\end{align}
The right hand side is an integrated version of the solution to the homogeneous Boltzmann equation
\begin{align}\label{eq:phononBE}
 &
   \partial_t W(k;t) = \lambda_3^2 \mathcal{C}_3[W(\cdot;t)](k) + \lambda_4^2 \mathcal{C}_4[W(\cdot;t)](k)
\, ,
\end{align}
where, using shorthand notations $\omega_\ell = \omega(k_\ell)$, $W_\ell = W(k_\ell)$, $\ell=0,1,2,3,4$,
\begin{align}\label{eq:defC3}
 &
  \mathcal{C}_3[W](k_0) =
  2^{\frac{7}{2}} \pi \ombasic^{-3} \sum_{\sigma\in \set{\pm 1}^2}
 \int_{\T^2}\! \rmd k\, 
 \delta(k_0+\sigma_1 k_1+\sigma_2 k_2) \delta(\omega_0+\sigma_1\omega_1+\sigma_2\omega_2)
 \nonumber \\ & \quad \times
 \prod_{\ell=0}^2 |\sin(\pi k_\ell)| \times
 \left(W_1 W_2+ \sigma_1 W_0 W_2+\sigma_2 W_0 W_1\right)
\, ,
\end{align}
and
\begin{align}\label{eq:defC4}
 &
  \mathcal{C}_4[W](k_0) =
 48 \pi \ombasic^{-4}
   \sum_{\sigma\in \set{\pm 1}^3} 
 \int_{\T^3}\! \rmd k\,
 \delta(k_0+\sum_{i=1}^3 \sigma_i k_i) \delta(\omega_0+\sum_{i=1}^3 \sigma_i \omega_i)
 \prod_{\ell=0}^3 |\sin(\pi k_\ell)|
 \nonumber \\ & \quad \times
 \left(W_1 W_2 W_3+ \sigma_1 W_0 W_2 W_3+\sigma_2 W_0 W_1 W_3+ \sigma_3 W_0 W_1 W_2\right)
\, .
\end{align}

Other interactions can be treated similarly.  For instance, the onsite perturbations 
have $\FT{\alpha}_3(k_1,k_2)= \lambda_3$ and $\FT{\alpha}_4(k_1,k_2,k_3)= \lambda_4$
and, for general dispersion relations $\omega$, yield the collision operators
\begin{align}\label{eq:defC3onsite}
 &
  \mathcal{C}_3[W](k_0) =
  4 \pi  \sum_{\sigma\in \set{\pm 1}^2}
 \int_{\T^2}\! \rmd k\, 
 \delta(k_0+\sigma_1 k_1+\sigma_2 k_2) \delta(\omega_0+\sigma_1\omega_1+\sigma_2\omega_2)
 \prod_{\ell=0}^3 \frac{1}{2 \omega_\ell}
 \nonumber \\ & \quad \times
 \left(W_1 W_2+ \sigma_1 W_0 W_2+\sigma_2 W_0 W_1\right)
\, ,
\end{align}
and
\begin{align}\label{eq:defC4onsite}
 &
  \mathcal{C}_4[W](k_0) =
 12 \pi 
   \sum_{\sigma\in \set{\pm 1}^3} 
 \int_{\T^3}\! \rmd k\,
 \delta(k_0+\sum_{i=1}^3 \sigma_i k_i) \delta(\omega_0+\sum_{i=1}^3 \sigma_i \omega_i)
 \prod_{\ell=0}^3 \frac{1}{2 \omega_\ell}
 \nonumber \\ & \quad \times
 \left(W_1 W_2 W_3+ \sigma_1 W_0 W_2 W_3+\sigma_2 W_0 W_1 W_3+ \sigma_3 W_0 W_1 W_2\right)
\, .
\end{align}
Detailed derivations of these onsite collision operators 
using standard perturbation expansions can also be found in the literature:
the third order collision operator in \cite{spohn05}
and the fourth order collision operator in \cite{ALS06}.

\subsection{Solution of the collisional constraints}

The above derivation of the Boltzmann collision operator is not mathematically rigorous.  In fact, the argument
used for neglecting the higher order terms and replacing ``$t-s$'' by ``$\infty$'' in the derivation of the energy conservation $\delta$-function
are at present mathematically uncontrollable approximations.  Here we assume that we are working in a regime in which these terms indeed 
can be neglected.  Hence, instead of trying to find out under
which conditions this would work for our particle chains, we start from the homogeneous Boltzmann equation
and proceed to study its solutions to obtain predictions which can later be compared with other results on the chains, such as computer simulations.

The weakest part of the argument lies in the rule that ``any integral over one of the $k$-components of the
rapidly oscillating factor $\rme^{-\ci t \Omega_n(k)}$ leads to integrable decay in $t$ and hence can be neglected''.  
This is particularly suspect if the integration is performed over a one-dimensional space: typically,
even the best decay estimates over a $d$-dimensional torus yield 
$\int_{\T^d}\!\rmd k\, \rme^{-\ci t\Omega(k)}=O(t^{-d/2})$ which is not integrable if $d=1$.
The decay of such oscillatory integrals is closely linked to the behaviour of the frequency function $\Omega_n$ around its zero set;
consider, for example, the explicitly integrable example $\int_0^t\!\rmd s\, \int \rmd k\, \rme^{-\ci s\Omega_n(k)}
= \int \rmd k\, \frac{\ci}{\Omega_n(k)} (\rme^{-\ci t\Omega_n(k)}-1)$ which is uniformly bounded in $t$, if $|\Omega_n(k)|^{-1}$ is integrable,
but which can blow up as fast as $O(t)$, for instance, if the set of $k$ with $\Omega_n(k)=0$ has a non-zero measure.

These subtleties are also reflected in the Boltzmann collision operator: the 
two $\delta$-functions need to be carefully integrated over before one can use the equation.
In the well-studied continuum setup of Boltzmann equations, the dispersion
relation is given by $\frac{1}{2} k^2$, $k\in \R^d$, and the integral contains 
$\delta(k_0+k_1-k_2-k_3)\delta((k_0^2+k_1^2-k_2^2-k_3^2)/2)$ which can be explicitly integrated
over, yielding the ``standard'' rarefied gas collision operator.

For a lattice system, the solutions to
$\Omega_n(k)=0$ are not so easy to handle.  As we will show below, there are however 
explicit solutions for the one-dimensional nearest neighbour dispersion relation, relevant to the 
FPU chains.  After the solution manifold has been found, one still needs to choose some parametrization
of the manifold to integrate over the energy conservation $\delta$-function, since this will
yield an additional factor to the remaining integrand.

In fact, even at this part of the procedure, some care is needed 
since the \defem{trivial solutions} to the collisional constraints, to be discussed later,
do not contribute to the total collision operator but would result in infinite factors if integrated over
using the above rule.  
Hence, the trivial solutions need to be removed from the solution manifold before integrating
out the collisional constraint $\delta$-functions.

We begin with a result which shows that nearest neighbour dispersion relations
are, in fact, somewhat pathological: they suppress all collisions involving three phonons.

\subsubsection{$\mathcal{C}_3=0$ for nearest neighbour dispersion relations}\label{sec:C3iszero}

Consider the nearest neighbour dispersion relation,
$\omega(k) =(1-2\delta \cos(2\pi k))^{1/2}$ with $0< \delta\le \frac{1}{2}$ (we ignore the overall prefactor here, since it does not affect the
solution manifold).  
To estimate the collision energy $\Omega_3 = \omega_0 + \sigma_1 \omega_1 + \sigma_2 \omega_2$, 
consider the following parametrization of $\omega$ as the magnitude of a complex number: 
Since $0<2 \delta\le 1$, we can define $r=\frac{1}{2}\mathrm{arcosh} (2\delta)^{-1}\ge 0$.  Then by explicit computation
\begin{align}
 \omega(k)=|z(k)|, \quad\text{for}\quad 
 z(k;\delta) = \sqrt{\delta} (\rme^{r} - \rme^{-r-\ci 2\pi k}) \, .
\end{align}
The above parametrization and the triangle inequality 
yield an upper bound for differences of dispersion relations. Namely, for $k,q\in \T$,
\begin{align}\label{eq:preb1}
 & |\omega(k)-\omega(k+q)|= \left|\,|z(k)|-|z(k+q)|\,\right|
%   \nonumber \\ & \quad
 \le |z(k)-z(k+q)| = \rme^{-r}\sqrt{\delta} |1- \rme^{-\ci 2\pi q}|\, .
\end{align}
Since $|1- \rme^{-\ci 2\pi q}|^2=2 -2 \cos( 2\pi q)=\delta^{-1}(2\delta - 2\delta\cos( 2\pi q))\le \delta^{-1} \omega(q)^2$, then
\begin{align}\label{eq:diffomegabound}
 & |\omega(k)-\omega(k+q)|\le  \rme^{-r}\omega(q)\, .
\end{align}

If $\delta< \frac{1}{2}$, we have $r>0$, and hence $\rme^{-r}<1$.
In addition, then $\omega(k)\ge c_\delta$, with $c_\delta=(1-2\delta )^{1/2}>0$.  
Thus if the momentum constraint $k_0+\sigma_1 k_1+\sigma_2 k_2=0$ holds, we have 
\begin{alignat}{2}
& \Omega_3= \omega_0 + \omega_1 + \omega_2 \ge 3 c_\delta\, , && \qquad\text{for}\quad\sigma_1=1=\sigma_2 \, ,\\
& \Omega_3= \omega_0 - \omega_1 + \omega_2 \ge (1-\rme^{-r}) c_\delta \, , && \qquad\text{for}\quad-\sigma_1=1=\sigma_2 \, ,\\
& \Omega_3= \omega_0 + \omega_1 - \omega_2 \ge (1-\rme^{-r}) c_\delta \, , && \qquad\text{for}\quad \sigma_1=1=-\sigma_2 \, ,\\
& \Omega_3= \omega_0 - \omega_1 - \omega_2 \le -(1-\rme^{-r}) c_\delta\, , && \qquad\text{for}\quad \sigma_1=-1=\sigma_2 \, .
\end{alignat}
Therefore, with a nearest neighbour interactions and pinning, the momentum constraint keeps the
three phonon energy well separated from zero, at least a distance $(1-\rme^{-r}) c_\delta$ apart.
In such a case, one should set $\mathcal{C}_3=0$ in the phonon Boltzmann equation.

If $\delta= \frac{1}{2}$, such as in the FPU models, 
we have $r=0$ and $c_\delta=0$, and equation (\ref{eq:diffomegabound}) does not rule out the existence of solutions to both 
constraints.  However, it is possible to improve the bound (\ref{eq:preb1}), by recalling that 
an equality in the triangle inequality holds if and only if one of the numbers is a \defem{non-negative} multiple of the other one.
This requires that for the given $k,q$ there should be $R\ge 0$ such that 
$(1-\rme^{-\ci 2\pi k_1})= R (1-\rme^{-\ci 2\pi k_2})$, where $k_1=k$, $k_2=k+q$, or vice versa.  For instance by a geometric argument,
it is straightforward to see that this happens if and only if $k_1=0$ or $k_2=0$.

As above, by inspecting the four sign combinations, we thus find that $|\Omega_3|>0$ unless $k_0=0$, $k_1=0$,
or $k_2=0$.  If $k_0=0$, we have $\Omega_3=0$ identically for $\sigma_1=-\sigma_2$, and only at the point $k_1=0=k_2$, if $\sigma_1=\sigma_2$.
If $k_0\ne 0$, we have $\Omega_3>0$ for $\sigma_1=1=\sigma_2$, and $\Omega_3=0$, only if one of the following conditions is satisfied:
% (1) $-\sigma_1=1=\sigma_2$, and $k_2=0$, (2) $\sigma_1=1=-\sigma_2$, and $k_1=0$, or (3) $\sigma_1=-1=\sigma_2$, and $k_1=0$ or $k_2=0$.
\begin{enumerate}
 \item $-\sigma_1=1=\sigma_2$, and $k_2=0$,
 \item $\sigma_1=1=-\sigma_2$, and $k_1=0$,
 \item $\sigma_1=-1=\sigma_2$, and $k_1=0$ or $k_2=0$.
\end{enumerate}
However, these solutions do not contribute to the collision operator in the FPU models as the collision 
operator $\mathcal{C}_3$ defined in (\ref{eq:defC3}) contains a prefactor $\prod_{\ell=0}^2 |\sin(\pi k_\ell)|$ 
which is zero on any of the above solution manifolds.  Therefore, for the FPU-chains, the lowest order 
kinetic theory would imply using $\mathcal{C}_3=0$.

The above argument straightforwardly generalizes to higher dimensions. As shown in Appendix 18.1 of \cite{spohn05},
also higher dimensional nearest neighbour dispersion relations with pinning have no
solutions to the constraints in the three-phonon collision operator and, hence, $\mathcal{C}_3=0$.  However, 
this property depends on the dispersion relation.  For instance, consider the
next-to-nearest neighbour interaction with $\FT{\alpha}(k)=(1-\cos(2\pi k))^2$.
Then the dispersion relation is $\omega(k)=1-\cos(2\pi k)$ and for $\sigma_1=1=-\sigma_2$
we thus have 
\begin{align}
 \Omega_3= \omega_0 + \omega_1 - \omega_2 = 1-\cos(2\pi k_0)-\cos(2\pi k_1)+\cos(2\pi (k_0+k_1))\, ,
\end{align}
which is zero if $k_1=\frac{1}{2}-k_0$.  This introduces 
solutions to the three-phonon collisional constraints in the next-to-nearest neighbour case.

\subsubsection{$\mathcal{C}_4$ for nearest neighbour dispersion relations}
\label{sec:C4gen}

From now on, we only consider the nearest neighbour dispersion relations.
We do not claim that this case is representative of the general case:
as the previous examples indicate, other types of behaviour might appear
for other dispersion relations.

Consider thus the collisional constraints in the four-phonon collision operator,
assuming a nearest neighbour dispersion relation.
If $\sigma_i=1$ for all $i$, by the above estimates, we have $\Omega_4>0$, unless
$\delta=\frac{1}{2}$ and $k_i=0$ for every $i=0,1,2,3$.

If $\sigma_i=-1$ for all $i$, we resolve the 
momentum constraint by integrating out $k_3$ which yields $k_3=k_0-k_1-k_2$.
Then 
\begin{align}
  \Omega_4 & = 
 \omega_0 - \omega_1 - \omega_2 -\omega_3 
    \nonumber \\ & 
 = \Omega_3(k_0,k_1,k_0-k_1; 1,-1,-1)
 +\Omega_3(k_0-k_1,k_2,k_0-k_1-k_2; 1,-1,-1) \, .
\end{align}
By the estimates derived in section \ref{sec:C3iszero}, we then have 
$\Omega_4<0$ uniformly in the case with pinning.
In addition, if $\delta=\frac{1}{2}$,  the only way to get $\Omega_4=0$
is that both of the terms above are zero, which happens
only if $k_1=0$ or $k_1=k_0$, {\em and\/} $k_2=0$ or $k_2=k_0-k_1$, i.e., $k_3=0$.
This implies that the only solutions to $ \Omega_4=0$ are the \defem{trivial solutions}
where two of the wave numbers $k_1,k_2,k_3$ are zero and the remaining one is equal to $k_0$.

The final degenerate case is found if $\sum_{i=1}^3 \sigma_i=1$.  
Then only one of $\sigma_i$ is negative; for notational simplicity 
let us suppose it is $\sigma_1$
(the other cases can then be obtained by permutation of the indices).
The momentum conservation implies then that $k_1-k_0=k_2+k_3$,
and hence 
\begin{align}
  \Omega_4 & = 
 \omega_0 - \omega_1 + \omega_2 + \omega_3 
    \nonumber \\ & 
 = \Omega_3(k_0,k_1,k_0-k_1; 1,-1,1)
 +\Omega_3(k_2,k_3,k_2+k_3; 1,1,-1) \, .
\end{align}
Thus $\Omega_4=0$ only for the trivial solution in the unpinned case,
namely, if $k_1=k_0$ and $k_2=0=k_3$.

The remaining sign-combinations have $1+\sum_{i=1}^3 \sigma_i=0$, i.e.,
the polarization of phonons is preserved in the collision.  These are also called 
collisions which \defem{conserve the phonon number}.
From the above results we can already conclude that every term in the collision
operator which is not phonon number conserving has only trivial solutions
to the energy constraint, and these solutions all require that two of the $k_i$,
$i=1,2,3$, are equal to zero and the last one is equal to $k_0$.

In particular, there are no trivial solutions in the pinned case, so these can again be safely neglected.
The same holds for the FPU chains, due to the prefactor $\prod_{\ell=0}^3 |\sin(\pi k_\ell)|$,
and since the solution involves a two-dimensional integral and
an integrand which has only a point-singularity.  To have more convincing an argument, one
should choose a regularization of the $\delta$-function and to show that the contribution from the 
resulting ordinary integrals vanishes as the regularization is removed.  An example of such 
a procedure is given in \cite[Section 3]{ls07} where it is used for showing that the trivial solutions 
in the number conserving case do not contribute to the linearized collision operator.

Here we are interested in the collision operators defined in (\ref{eq:defC4}) and (\ref{eq:defC4onsite}).
Using the permutation properties of the integrands multiplying the constraint $\delta$-functions,
we thus arrive at the following simplified forms for the collision operators:
in the FPU chains, the standard kinetic argument yields
\begin{align}\label{eq:defCtot}
 &
  \mathcal{C}[W](k_0) =
 9 \pi\, (2 \lambda_4)^2 \ombasic^{-8}
 \int_{\T^3}\! \rmd k\,
\delta(k_0+k_1-k_2-k_3) \delta(\omega_0+\omega_1-\omega_2-\omega_3)
 \prod_{\ell=0}^3 \omega_\ell
 \nonumber \\ & \quad \times
 \left(W_1 W_2 W_3+ W_0 W_2 W_3-W_0 W_1 W_3-W_0 W_1 W_2\right)
\, ,
\end{align}
and for onsite nonlinear interactions 
\begin{align}\label{eq:defCtotonsite}
 &
  \mathcal{C}[W](k_0) =
 \frac{9 \pi}{4} \lambda_4^2
 \int_{\T^3}\! \rmd k\,
 \delta(k_0+k_1-k_2-k_3) \delta(\omega_0+\omega_1-\omega_2-\omega_3)
 \prod_{\ell=0}^3 \omega_\ell^{-1}
 \nonumber \\ & \quad \times
 \left(W_1 W_2 W_3+ W_0 W_2 W_3-W_0 W_1 W_3-W_0 W_1 W_2\right)
\, .
\end{align}

The two operators are nearly identical, differing only by 
the powers of the $\omega(k_\ell)$ factors.  This difference however has an important
consequence to the properties of the solutions to these two equations:
the onsite equation (\ref{eq:defCtotonsite}) predicts finite thermal conductivity from the Green--Kubo formula, whereas
the FPU chain equation predicts an infinite result.  Qualitatively, one can 
understand this by noticing that in the unpinned case $\omega$ has zeroes which
leads to enhancement of collisions for the onsite anharmonicity, but which suppresses them in the FPU case.
However, the qualitative argument is not sufficient to determine the magnitude of the effect, and we
need to do a more careful study of the properties of the solutions to obtain predictions
about the thermal conductivities.

\subsubsection{Integration of the collisional constraints in FPU chains}\label{sec:FPUconst}

Let us begin with the unpinned nearest neighbour models relevant to the FPU chains.
Since then $\omega(k)=\sqrt{2} \ombasic |\sin(\pi k)|$, it is analytically simpler 
to reparametrize the wave-number integrals by changing variables from $k\in \T$
to $p=2\pi k$ which belongs to the half-open interval $I = [0,2\pi)$.   Then 
$\omega(k)$ is replaced by $\widetilde{\omega}(p)=\sqrt{2} \sin(p/2)$ where the absolute value is 
not needed since it was possible to restrict the values of $p$ to an interval where $\sin(p/2)$ is positive.

Although very useful for derivation of explicit parametrizations of the solution manifold, it should be stressed
that some care is needed with this choice: \defem{it is crucial below that all arithmetic involving $p\in I$
is performed modulo $I$}.  For instance, if $p_1=\pi/2$ and $p_2 = 3\pi/2$, we have to use 
$p_1-p_2=+\pi$ in $\widetilde{\omega}(p_1-p_2)$
to get its value correctly.  This rule of ``modulo $I$ arithmetic'' will be applied without further mention below.

We make the change of variables also in the spatially homogeneous Wigner function and consider $\tilde{W}_t(p) = W_t(p/(2\pi))$.
Let us for simplicity 
drop the tilde from here and from the reparametrized dispersion relation, and denote them simply by $W_t(p)$ and $\omega(p)$.
With these conventions, the phonon Boltzmann equation of the FPU chain becomes
$\partial_t W_t(p) = \mathcal{C}[W_t](p)$ where
\begin{align}\label{eq:defFPUBE}
% & \partial_t W_t(p) = \mathcal{C}[W_t](p),\quad \text{where}\\
 & \mathcal{C}[W](p_0) =
 \frac{9}{\pi}\lambda_4^2 \ombasic^{-8}
 \int_{I^2}\! \rmd p_1\rmd p_2\, \delta(\Omega(p))
 \prod_{\ell=0}^3 \omega_\ell
 \nonumber \\ & \quad \times
 \left(W_1 W_2 W_3+ W_0 W_2 W_3-W_0 W_1 W_3-W_0 W_1 W_2\right)
\, ,
\end{align}
with $\omega_\ell = \omega(p_\ell)$, $ W_\ell = W(p_\ell)$, and 
\begin{align}
&  \Omega(p) = \omega_0 + \omega_1 - \omega_2 -\omega_3\, ,\quad 
p_3 = (p_0+p_1-p_2) \bmod I\, .\label{eq:FPUconstraints} 
\end{align}

The detailed solution of the collisional constraints, $\Omega(p)=0$ in (\ref{eq:FPUconstraints}), can be found in \cite{ls07}.
By Corollary 3.3. there, we can conclude that both constraints are satisfied exactly for those $p_\ell$, $\ell=0,1,2,3$,
which satisfy one of the following three relations
\begin{enumerate}
 \item $p_2=p_0$,
 \item $p_1=p_2$, or
 \item $p_1=h(p_0,p_2)\bmod I$, where $h$ is defined using the standard, non-periodic, arithmetic in
 \begin{align}\label{eq:FPUnontrivsol}
  h(x,y) = \frac{y-x}{2} +
  2 \arcsin \! \left( \tan \frac{|y-x|}{4} \cos \frac{y+x}{4} \right)
\end{align}
with $\arcsin$ denoting the principal branch with values in 
$[-\pi/2,\pi/2]$.
\end{enumerate}
The solutions satisfying item 1 or 2 are called \defem{perturbative} or \defem{trivial}, while the solution 
satisfying 3 is called \defem{non-perturbative}.  The nomenclature can be understood by expanding the constraint
$\Omega$ around small values of $p_\ell\in \R$ using $\omega(p) \approx 2^{-\frac{1}{2}} |p|(1-p^2/24)$.  The resulting
equation then has only 1 and 2 as its solutions.

The remaining energy conservation $\delta$-function can then
be formally resolved by integrating over a suitably chosen direction in the $p_1,p_2$-variables.
For instance, choosing the $p_1$-integral for this purpose
would yield for any $p_2\ne p_0$ and for any continuous periodic function $G$,
\begin{align}\label{eq:exampleOmegaint}
& \int_I\! \rmd p_1\, \delta(\Omega(p)) G(p_0,p_1,p_2) 
\nonumber \\ & \ 
= \frac{1}{|\partial_2\Omega(p_0,p_2,p_2)|} G(p_0,p_2,p_2)
+ \frac{1}{|\partial_2\Omega(p_0,h(p_0,p_2),p_2)|} G(p_0,h(p_0,p_2),p_2).
\end{align}
However, this procedure needs to be used with some care: for instance, in the present case
$\int_I\! \rmd p_2\, |\partial_2\Omega(p_0,p_2,p_2)|^{-1}=\infty$, 
and thus the first term on the right hand side of (\ref{eq:exampleOmegaint}) would typically 
diverge when integrated over $p_2$.  

This problem is resolved in the collision operator by the alternating signs in its integrand which
guarantee that the integrand vanishes at the trivial solutions.  This can even be proven rigorously
if $1/W$ is sufficiently regular, say twice continuously differentiable, by using the following equivalent form 
for the collision operator:
\begin{align}%\label{eq:defFPUBE}
% & \partial_t W_t(p) = \mathcal{C}[W_t](p),\quad \text{where}\\
 & %\mathcal{C}[W](p_0) =
 \frac{9}{\pi}\lambda_4^2 \ombasic^{-8}
 \int_{I^2}\! \rmd p_1\rmd p_2\, \delta(\Omega(p))
 \prod_{\ell=0}^3 (\omega_\ell W_\ell)
% \nonumber \\ & \quad \times
 \left(W^{-1}_0 + W^{-1}_1 -W^{-1}_2-W^{-1}_3\right)
\, .
\end{align}
However, we do not wish to go into more detail here, but just add the contribution from the trivial solutions
to the class of terms which are assumed to be negligible in the kinetic theory of FPU lattices.

If the trivial solutions are neglected, we can use results from \cite{ls07}, which rely on the
explicit form of the non-trivial solution $h$, and obtain the fully integrated form
\begin{align}%\label{eq:defFPUBE}
 & \mathcal{C}[W](p_0) =
 \frac{9 \sqrt{2}}{\pi}\lambda_4^2 \ombasic^{-8}
 \int_{0}^{2\pi}\!\rmd p_2\, \frac{1}{\sqrt{F_+(p_0,p_2)}}
 \prod_{\ell=0}^3 \omega_\ell
 \nonumber \\ & \quad \times
 \left(W_1 W_2 W_3+ W_0 W_2 W_3-W_0 W_1 W_3-W_0 W_1 W_2\right)
\, ,
\end{align}
where $p_1=h(p_0,p_2)$, $p_3=p_0+p_1-p_2$, and
\begin{align}\label{eq:defFpm}
F_{\pm}(x,y) = \left(\cos \frac{x}{2} + \cos \frac{y}{2}\right)^2 \pm 
    4 \sin \frac{x}{2} \sin \frac{y}{2}  \, .
\end{align}
(See for instance, Lemma 3.4 in \cite{ls07} for more details.  From the Lemma 
a weak convergence of the integrals with a regularized $\delta$-function 
$\delta_\epsilon(\Omega) = \epsilon \pi^{-1} (\epsilon^2 + \Omega^2)^{-1}$, $\epsilon>0$,
and assuming continuity of $W$, can be established.)

The function $F_-$ defined in (\ref{eq:defFpm}) is related to the change of variables which corresponds to using 
$p_2$ instead of $p_1$ to integrate out the energy $\delta$-function.  Namely, as proven in Lemma 3.5 
in \cite{ls07}, 
\begin{align}\label{eq:lemmaK2toK1}
& \int_{I^2}\rmd p_0 \rmd p_2 \, \frac{1}{\sqrt{F_+(p_0,p_2)}} G(p_0,h(p_0,p_2))
% \nonumber \\ & \quad 
= 2 \int_{I^2}\rmd p_0\rmd p_1\, \frac{\cf(F_-(p_0,p_1)>0)}{\sqrt{F_-(p_0,p_1)}} G(p_0,p_1) \, ,
\end{align}
for any $G$ for which either of the two integrals converges.  The characteristic function restricts the
integral to the subset of $I^2$ in which the argument of the square root is positive.  Let us also mention that the
change of variables becomes more
involved if $G$ also depends on $p_2$ directly since each pair $(p_0,p_1)$, for which $F_-(p_0,p_1)>0$,
has two distinct values $p_2$ solving the energy constraint.
Further details about these solutions can be found from the proof of the above mentioned Lemma 3.5.

\subsubsection{Integration of the collisional constraints for onsite nonlinearity}
\label{sec:solOnsite}

To allow a comparison, let us also consider the case with an onsite nonlinearity.
If the harmonic part has no pinning, then the above FPU discussion applies immediately,
since these models have the same dispersion relation and differ only by the factors in the 
integrand.  We thus find the following integrated form of the collision operator,
if $\delta=\frac{1}{2}$ in the general nearest neighbour dispersion relation:
\begin{align}%\label{eq:defFPUBE}
 & \mathcal{C}[W](p_0) =
 \frac{9}{2^{7/2}\pi}\lambda_4^2
 \int_{0}^{2\pi}\!\rmd p_2\, \frac{1}{\sqrt{F_+(p_0,p_2)}}
 \prod_{\ell=0}^3 \omega_\ell^{-1}
 \nonumber \\ & \quad \times
 \left(W_1 W_2 W_3+ W_0 W_2 W_3-W_0 W_1 W_3-W_0 W_1 W_2\right)
\, ,
\end{align}
where $p_1=h(p_0,p_2)$, $p_3=p_0+p_1-p_2$, and $F_+$ has been defined in (\ref{eq:defFpm}).
(We continue to neglect contributions from the trivial solutions.)

The analytic structure of the solution manifold 
gets more complicated once the dispersion relation has pinning, i.e.,
when $\delta<\frac{1}{2}$.  Nevertheless, it is still possible to find
a function $h(p_0,p_2)$ such that the non-perturbative solution is 
parametrized by a condition
\begin{align}
p_1 = h( p_0,p_2;\delta)\, .
\end{align}
In other words, the enumeration by the three conditions in the previous section
continues to hold, only with a function $h$ which depends on $\delta$.
Naturally, $h( p_0,p_2;\frac{1}{2})$ is then given by (\ref{eq:FPUnontrivsol}).

Since to our knowledge the explicit form of the solution function $h$ is not available
in the literature, let us present some details for its derivation.
For the moment, let us return to standard arithmetic for $p_\ell$ and
consider $p_i\in \R$, $i=0,1,2$, and take $p_3=p_0+p_1-p_2$.
Next change variables from $p_\ell$ to
\begin{align}
 & u = \frac{p_2-p_0}{2} = \frac{p_1-p_3}{2}\, ,\quad
 v = \frac{p_2+p_0}{2}\, ,\quad
 w = \frac{p_1+p_3}{2}\, .
\end{align}
Then
\begin{align}
 p_1 = w+u\, ,\quad
 p_3 = w-u\, ,\quad
 p_2 = v+u\, ,\quad
 p_0 = v-u\, .
\end{align}
Therefore, the energy constraint $\Omega_4=0$ is equivalent to
\begin{align}\label{eq:OMtog}
 &g(w,u) = g(v,u)\, ,
\end{align}
where, for simplicity, we set $\ombasic=1$ and then define
\begin{align}
g(v,u) =\sqrt{1-2\delta \cos(v+u)}-\sqrt{1-2\delta \cos(v-u)}= \omega_2-\omega_0\, ,
\end{align}
and thus $g(w,u)=\omega_1-\omega_3$.  

Since $g(v,\pi n)=0$ for all $v$ and $n\in \Z$, (\ref{eq:OMtog}) is solved by any $v,w\in \R$ if $u\in \pi \Z$.
This corresponds to the trivial solution $p_2=p_0 \bmod 2\pi$.

Assume thus $u\not\in \pi \Z$.  To fix a sign convention, let us next suppose that
$w,v\in J=(-\pi,\pi]$ which can always be achieved without changing $u$ by shifting $p_\ell$ by
a suitably chosen integer multiple of $2\pi$.
Since
\begin{align}
 g(v,u) = 2\delta \frac{\cos(v-u)-\cos(v+u)}{\omega_2+\omega_0}
 = 4\delta \frac{\sin u \sin v}{\omega_2+\omega_0},
\end{align}
where $\sin u \ne 0$ and $\omega_\ell \ge 0$, (\ref{eq:OMtog}) can only hold if $\sin v$ and $\sin w$ have the same sign.
For $w,v\in J$ this is equivalent to requiring $\sign(w)=\sign(v)$.

To solve (\ref{eq:OMtog}), we first note that 
\begin{align}
 g(v,u)^2 & = 2(1- 2\delta \cos u \cos v - \omega_0\omega_2)\, ,\\
  g(w,u)^2 &= 2(1- 2\delta \cos u \cos w - \omega_1\omega_3)\, .
\end{align}
Hence, if (\ref{eq:OMtog}) is true, we need to have
\begin{align}
 \omega_1\omega_3 = -2\delta \cos u (\cos w -\cos v)+  \omega_0\omega_2\, .
\end{align}
We square both sides one more time, and use the identities
\begin{align}
 & \omega_1^2\omega_3^2 = 1 - 4 \delta \cos u \cos w  + 4 \delta^2 \cos^2 w - 4 \delta^2 \sin^2 u\, ,\\
 & \omega_0^2\omega_2^2 = 1 - 4 \delta \cos u \cos v  + 4 \delta^2 \cos^2 v - 4 \delta^2 \sin^2 u\, ,
\end{align}
which follow from the trigonometric relation $\cos(a+b)\cos(a-b)= \cos^2 a - \sin^2 b$.

The result can be written in terms of the variable $y=\cos w-\cos v$, and we find that  (\ref{eq:OMtog})
implies the equation
\begin{align}
 y^2 4 \delta^2 \sin^2 u   - y 4 \delta (\cos u (1-\omega_0\omega_2)-2\delta \cos v ) = 0\, .
\end{align}
Since $\sin u\ne 0$, this equation has exactly two solutions for $y$.  The solution $y=0$
has $\cos w=\cos v$ which yields only the solution $w=v$ for (\ref{eq:OMtog}) with $w, v\in J$
since then $w$ and $v$  must have the same sign.   But then $p_2=p_1$, so $y=0$ corresponds to
the second trivial solution.

This leaves only the following candidate for a non-perturbative solution:
\begin{align}\label{eq:ynontriv}
 y = \frac{\cos u (1-\omega_0\omega_2)-2\delta \cos v}{\delta \sin^2 u}\, .
\end{align}
The denominator appears to lead to singularities at $u=0$ and at $\delta=0$.
However, both singularities are removable.  Namely, (\ref{eq:ynontriv}) implies that
\begin{align}%\label{eq:ynontriv}
 & y+ 2 \cos v = \cos u
 \frac{1-2\delta \cos u \cos v-\omega_0\omega_2}{\delta \sin^2 u}
% \nonumber \\ & \quad
 = 
 \frac{4\delta \cos u \sin^2 v}{1-2\delta \cos u \cos v+\omega_0\omega_2}\, .
\end{align}
Since the left hand side is equal to $\cos w+\cos v$ and $w,v$ have the same sign,
we again get only one solution which expresses $w$ as a function of $u,v$.

In summary, the above computation yields the following expression for the
non-per\-tur\-ba\-tive solution.  Choose $p_0,p_2\in J=(-\pi,\pi]$.  Then also $v\in J$ and we
have $p_1=h(p_0,p_2;\delta) \bmod 2\pi$, where the non-perturbative solution is given by
\begin{align}\label{eq:defhnonptonsite}
   & h(p_0,p_2;\delta) = \frac{p_2-p_0}{2}
 +
 \sign\!\left(\frac{p_2+p_0}{2}\right) 
 \arccos 
  \biggl[ -\cos \frac{p_2+p_0}{2} 
 \nonumber \\ & \qquad
  +  2 \delta \frac{\sin p_0+\sin p_2}{1-\delta(\cos p_0+ \cos p_2)+\sqrt{(1-2\delta \cos p_0)(1-2\delta \cos p_2)}} \sin \frac{p_2+p_0}{2} \biggr]
\end{align}
with $\arccos \in [0,\pi]$ denoting the principal branch.  In addition, to have the correct value of $h$ at the apparent discontinuity 
$p_2=-p_0$, we also define $\sign(0)=1$: this yields $h(p_0,p_2)\to \pi-p_0=h(p_0,-p_0)$, as $p_2\searrow -p_0$, 
and $h(p_0,p_2)\to h(p_0,-p_0)-2\pi$, as $p_2\nearrow -p_0$.
We skip the rest of the details, namely proving that the above $\arccos$ is
always well-defined (i.e., its argument lies in $[-1,1]$) and that the result always provides a solution to the original constraint problem.

\begin{figure}[t]
  \begin{center}
    \includegraphics*[width=0.9\textwidth]{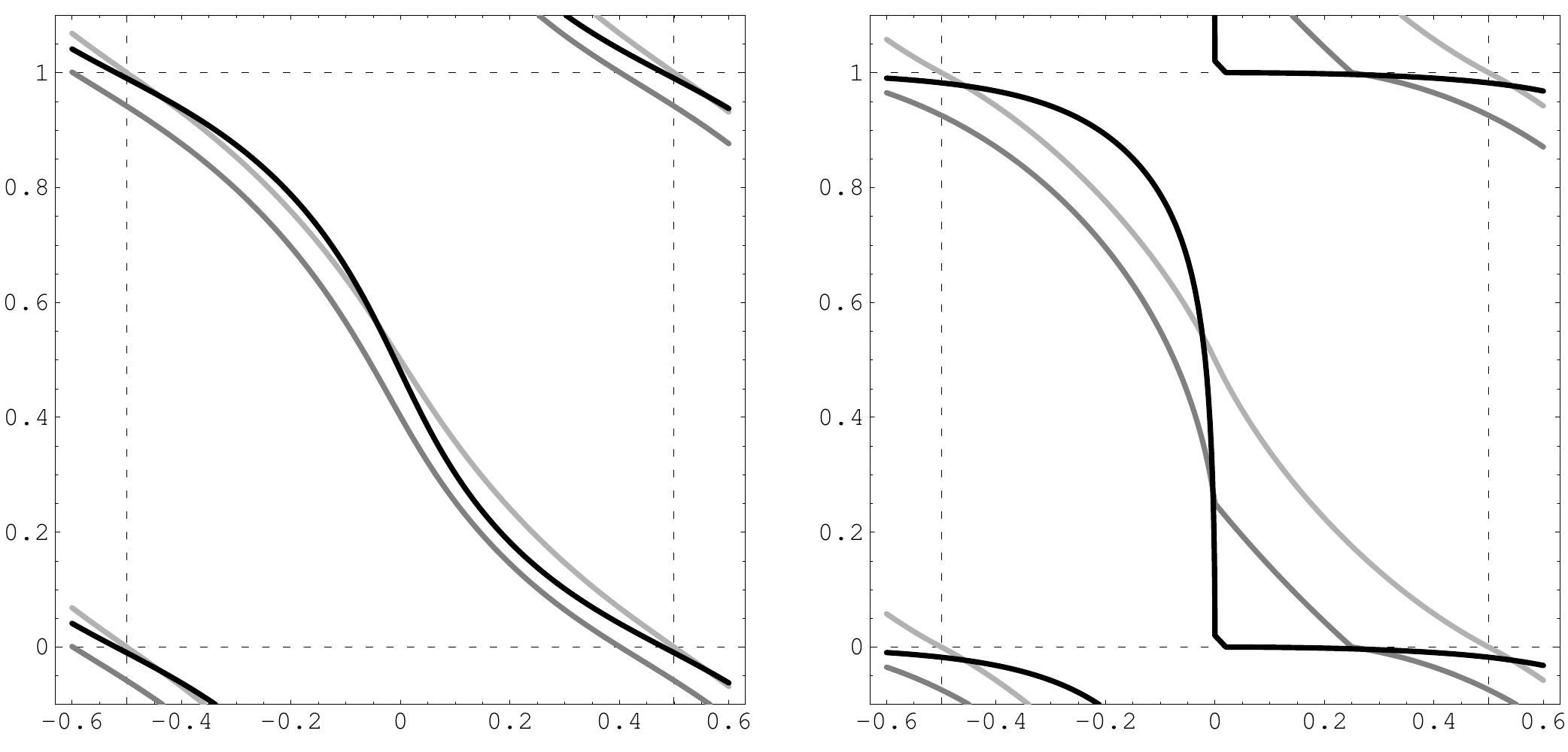}
    \caption{The non-perturbative solution $k_1=\frac{1}{2\pi}h(2\pi k_0,2\pi k_2;\delta)$
      as a function of $k_0$ for fixed $k_2$ and $\delta$.
      On the left, $\delta=0.4$, and on the right, $\delta=0.5$.
      For each $\delta$, three solutions are plotted, corresponding to
      $k_2=0.02$ (black), $0.25$ (dark grey), and $0.5$ (light grey).  We plot
      $\mathbb{T}\times \mathbb{T}$ in the extended
      zone scheme, the dashed lines are the
      boundaries of a unit cell. For $\delta < 0.5$ the non-perturbative
      solution is 
smooth, while for  $\delta = 0.5$ there are cusp singularities. (Reprinted from \cite{ALS06})}
    \label{fig:umklapps}
  \end{center}
\end{figure}

In general, for a fixed $p_2$ the function $h$ is
continuous, one-to-one, and satisfies
\begin{align}
\omega(p_0)+\omega(h(p_0,p_2))-\omega(p_2)-\omega(p_0+h(p_0,p_2)-p_2)=0\,. 
\end{align}
In Fig.~\ref{fig:umklapps}
we display a few non-perturbative solutions, at
$\delta=0.4$ and at $\delta=0.5$, for three different values of $p_2$.
For small $\delta$ one finds
\begin{align}
h(p_0,p_2;\delta)=\pi-p_0-\delta\big(\sin p_0
 +\sin p_2 \big)+\mathcal{O}(\delta^2)\,, 
\end{align}
which reasonably well approximates the left hand side of Fig.~\ref{fig:umklapps}.
The existence of the above limiting solution was first noted in \cite{lefev06} and it
served as an important motivation for looking closer at the predictions from
one-dimensional kinetic theory of phonons.

To resolve the collisional constraints, we integrate over $p_1$, as in the FPU case.
Since $\partial_{p_1} \Omega(p_0,p_1,p_2) = \omega'(p_1)-\omega'(p_0+p_1-p_2)$
and $\omega'(p)= \delta \sin p / \omega(p)$, this yields the collision operator
\begin{align}\label{eq:resolvedonsiteBE}
 & \mathcal{C}[W](p_0) =
 \frac{9}{16 \pi \delta}\lambda_4^2 
 \int_{-\pi}^{\pi}\!\rmd p_2\, 
 \frac{1}{\omega_0 \omega_2 |\omega_3 \sin p_1- \omega_1 \sin p_3|}
 \nonumber \\ & \quad \times
 \left(W_1 W_2 W_3+ W_0 W_2 W_3-W_0 W_1 W_3-W_0 W_1 W_2\right)
\, ,
\end{align}
where $p_1=h(p_0,p_2;\delta)$ and $p_3=p_1+p_0-p_2$.

\section{Energy transport in the kinetic theory of phonons}

\subsection{Entropy and H-theorem of the phonon Boltzmann equations}
\label{sec:Htheorem}

For ergodic systems, time averages will converge to ensemble averages.  We have been working under
the assumption that energy is the only ergodic variable for our particle chains.  In addition, for infinite volume systems
the canonical and microcanonical ensembles should agree, so to study the convergence towards equilibrium, one could
inspect the evolution of the standard entropy density. Since it is maximized for the given energy exactly by the canonical
ensemble, this should yield an increasing function, at least asymptotically for large times.

Consider thus a probability measure $\rho(q,p) \rmd^{N\!} q\,\rmd^{N\!} p$ for the state of the finite periodic 
system with $N$ particles.  The corresponding entropy
is defined by 
\begin{align}
 S[\rho] = -\int \rmd^{N\!} q\,\rmd^{N\!} p\, \rho(q,p) \ln \rho(q,p)\, .
\end{align}
For small couplings, the canonical Gibbs measure is approximately Gaussian.  
In general, for measures which are nearly Gaussian in 
the random vector $X$, and assuming that $X$ has a mean $\mu$ and a covariance matrix $C$,
we have $\ln \rho(q,p) \approx -\frac{1}{2}\left( \ln [\det (2\pi C)] + (X-\mu)^T C^{-1} (X-\mu)\right)$.
This implies that the entropy of such nearly Gaussian measures satisfies
\begin{align}
 S[\rho] \approx \frac{1}{2} \ln [\det (2\pi C)] + \frac{1}{2} \tr 1\, .
\end{align}

Thus the entropy per particle, $N^{-1} S[\rho]$, is then approximately equal to a constant plus
$\frac{1}{2 N} \ln (\det C)$.  For a spatially homogeneous particle chain, one has
$C=\mathcal{F}^{-1}\FT{C}\mathcal{F}$ where $\mathcal{F}$ denotes the Fourier-transform.  Therefore,
for spatially homogeneous nearly Gaussian states and large enough $N$
\begin{align}
  N^{-1} S[\rho] \approx \text{constant} + \frac{1}{2}\int\!\rmd k \ln \det F(k) \, ,
\end{align}
where $F(k)$ is a $2\times 2$ matrix defined by
\begin{align}
F(k)= \int \rmd k' \begin{pmatrix}
  \kappa[\FT{q}(k')^*,\FT{q}(k)] & \kappa[\FT{q}(k')^*,\FT{p}(k)]\\
  \kappa[\FT{q}(k')^*,\FT{p}(k)] & \kappa[\FT{p}(k')^*,\FT{p}(k)]
 \end{pmatrix}\, .
\end{align}
This can be expressed in terms of the phonon fields, and hence their Wigner function $W(k)$,
using the relations given in (\ref{eq:invphonon}):
\begin{align}
F(k)= \frac{1}{2} \begin{pmatrix}
  \omega(k)^{-1} (W(k)+W(-k)) & -\ci (W(k)-W(-k))\\
  \ci (W(k)-W(-k)) & \omega(k) (W(k)+W(-k))
 \end{pmatrix}\, .
\end{align}
Therefore, $\det F(k) = W(k) W(-k)$, and thus we obtain the following definition of entropy density
for phonon systems in a kinetic scaling limit
\begin{align}
 S[W] = \int\!\rmd k \ln W(k) \, .
\end{align}

After the above preliminaries, it is satisfying to find out that the above reasoning leads 
to an entropy functional which satisfies an H-theorem for phonon Boltzmann equations.
This holds quite generally but let us check it explicitly only for the
homogeneous Boltzmann equations $\partial_t W(k) = \mathcal{C}[W(k)]$ where 
the collision operator $\mathcal{C}$ is given either by (\ref{eq:defCtot}) or by (\ref{eq:defCtotonsite}).
In both cases, there is a positive function $G$ such that
\begin{align}\label{eq:Ctotform2}
 &
  \mathcal{C}[W](k_0) =
 \int_{\T^3}\! \rmd k\,
\delta(k_0+k_1-k_2-k_3) \delta(\omega_0+\omega_1-\omega_2-\omega_3)
\prod_{\ell=0}^3 \left[G(\omega_\ell) W_\ell\right]
 \nonumber \\ & \quad \times
 \left(W_0^{-1}+W_1^{-1}- W_2^{-1} -W_3^{-1}\right)
\, .
\end{align}
Since the integrand, including both constraints, is symmetric under $0\leftrightarrow 1$ and
$2\leftrightarrow 3$, and antisymmetric under $(0,1)\leftrightarrow (2,3)$, we find that
\begin{align}\label{eq:derHtheorem}
& \partial_t S[W_t] = \int\!\rmd k_0 W_0^{-1} \partial_t W_t(k_0)
 \nonumber \\ & \quad 
 = \frac{1}{4} \int_{\T^{\set{0,1,2,3}}}\! \rmd k\,
\delta(k_0+k_1-k_2-k_3) \delta(\omega_0+\omega_1-\omega_2-\omega_3)
 \nonumber \\ & \qquad \times
\prod_{\ell=0}^3 \left[G(\omega_\ell) W_\ell\right]
 \left(W_0^{-1}+W_1^{-1}- W_2^{-1} -W_3^{-1}\right)^2
 \ge 0
 \, .
\end{align}

Hence, the entropy $S[W_t]$ is increasing along solutions of the Boltzmann
equation, i.e., it satisfies an H-theorem.  The entropy production functional $D[W]$,
satisfying $\partial_t S[W_t] = D[W_t]$, also has an explicit
form which can be read off from the right hand side of (\ref{eq:derHtheorem}).

\subsection{Steady states}
\label{sec:BEsteadys}

The H-theorem (\ref{eq:derHtheorem}) allows to classify all steady states of the 
phonon Boltzmann equation.  Namely, suppose that $\bar{W}$ is a steady state,
i.e., $W_t=\bar{W}$ is a solution to the phonon Boltzmann equation.   Since then
the left hand side of (\ref{eq:derHtheorem}) is zero, the integrand defining the entropy
production has to vanish almost everywhere on the solution manifold.

We are only interested in nondegenerate steady states which correspond to
functions $W$ for which $W>0$ almost everywhere.  Then it is possible
to apply the previous explicit solutions $p_1=h$ of the collisional constraints
to simplify the above problem.  Namely, 
for any steady state $W$ its inverse $f(p)=W(p)^{-1}$ then
necessarily satisfies
\begin{align}\label{eq:steadyeq}
f (p_0)+f(h(p_0,p_2;\delta))-f(p_2)-f(p_0-p_2+h(p_0,p_2;\delta))=0\,,
\end{align}
for almost every $p_0$ and $p_2$.

Due to the collisional constraints, any linear combination 
$f(k) = \beta \omega(k) + \alpha$ is obviously
a solution to the functional equation 
(\ref{eq:steadyeq}).  
As shown in \cite{spohn05,spohn06},
these are quite generally the only solutions for two- and higher-dimensional 
crystals.

The one-dimensional case is more intricate.  For instance,
$h(p_0,p_2;\delta)\to \pi -p_0$ when $\delta\to 0$.  Hence, in the limit,
any function $f$ satisfying the symmetry requirement $f(p)=-f(\pi -p)$
is a solution to (\ref{eq:steadyeq}).
However, we do not expect that any new solutions would appear 
for the nearest neighbour dispersion relations, i.e., if $\delta>0$.
This has even been rigorously proven to be the case for the unpinned dispersion
relation with $\delta=\frac{1}{2}$;
see Section 5 in \cite{ls07} for details.

In summary, we expect that the only steady states for which $W$ is integrable, 
are given by 
\begin{align}\label{eq:defWeql}
 \Weql (k) = \frac{1}{\beta (\omega(k)-\mu)}\, ,
\end{align}
where $\beta>0$ and $\mu<\min \omega$.  The last condition is necessary,
since if there were a point $k_0$ for which
$\omega(k_0)=\mu$, this would lead to a nonintegrable singularity at $k_0$ for $W$
(the singularity $|x|^{-1}$ is not integrable around $x=0$ in one dimension).

Using (\ref{eq:Ctotform2}) it is straightforward
to check that indeed $\mathcal{C}[\Weql]=0$ for the functions defined in (\ref{eq:defWeql}).
Hence, all of them are true steady states of the phonon Boltzmann equation.
The conservation of phonon number is reflected in the appearance of the second parameter
$\mu$ for the steady states.  It is expected to be a spurious chemical potential, and
one expects that $\mu\to 0$ eventually as $t\to \infty$.  However,
this process is not captured by the standard kinetic approximation which we have applied here,
and resolving the issue would require a different approach.

% and most likely this part of the equilibration process takes place at longer than kinetic
% time scales, for $t=O(\lambda^{-2-p})$ for some $p>0$.

Let us also mention in passing that there are additional solutions to 
$\mathcal{C}[W]=0$ if one allows $W$ to be a distribution.
For instance, $W(k)=\delta(k)$ formally provides a solution to the FPU collision operator.
This is not surprising since the additional constraints given by such distributions
can restrict the solution manifold further.  In general, one should treat such distributional
solutions with great care since it is far from obvious that the 
oscillatory terms, which are neglected in the derivation
of the phonon Boltzmann equation, remain lower order contributions for such
non-chaotic initial data.  However, there are known examples where distributional solutions
seem to play a physical role,
such as in the kinetic theory of Bose-Einstein condensation: see \cite{lu14} and the references therein for more details.

\subsection{Green--Kubo formula and the linearized Boltzmann equation}
\label{sec:linearizedBE}

Let us now return to the Green--Kubo formula for thermal conductivity.
We consider here the kinetic prediction for the leading harmonic part of the current-current
correlation function, as derived in Section \ref{sec:GKformula}.
We choose the basic one-parameter canonical Gibbs measure at temperature $T>0$ 
as the initial steady state.
Its limiting covariance as $\lambda\to 0$ then has a Wigner function $\Weql$ 
as given in (\ref{eq:defWeql}) with $\beta=1/T$ and $\mu=0$.  (As discussed earlier, we expect the true
steady states to have $\mu=0$ even though the steady states of the kinetic model can have $\mu\ne 0$.)

In fact, the results in Sec.~\ref{sec:GKformula} imply that the evolution of the harmonic part of the
Green--Kubo correlator is determined by the phonon Boltzmann equation linearized around
its stationary solution $\Weql(k;\beta) = 1/(\beta \omega(k))$.
We begin with the following approximation for
the Green--Kubo correlator obtained from (\ref{eq:Ctharm2}) and (\ref{eq:vepev}):
\begin{align}
 C(t;\beta) \approx \ombasic^4 \delta^2 \int_\T \! \rmd k \, \phi_0(k) \partial_\vep W^{(\vep)}_t(k)|_{\vep=0} \, ,
\end{align}
where we have written the covariance in terms of the Wigner function $W^{(\vep)}_t(k)$ and defined
\begin{align}
 \phi_0(k) = \sin (2\pi k)\, .
\end{align}
The Wigner function is computed using
expectation $\left\langle \cdot \right\rangle^{(\vep)}$ over the stochastic
process which starts from a perturbation of the stationary state whose initial Wigner function 
converges to $W^{(\vep)}_0(k)=1/(\beta\omega(k)-\vep \phi_0(k))$ as $\lambda\to 0$.

Since 
$\vep |\phi_0(k)|\le 2\pi \vep |k|$, the Wigner function is positive for sufficiently small $\vep>0$, 
even in the FPU-models for which $\omega(0)=0$: this follows from the observation that it is always possible to find some
$c_0>0$ such that  $\omega(k)\ge c_0 |k|$ in a neighbourhood of $0$.
Hence, $h_t(k)=\partial_\vep W^{(\vep)}_t(k)|_{\vep=0}$ is well defined and its initial value is given by 
$h_0=(\Weql)^2 \phi_0$.  Its time-evolution can be determined using the equality
$\partial_t h_t(k) = \partial_\vep \partial_t W^{(\vep)}_t(k)|_{\vep=0}$ and the kinetic theory approximation 
$\partial_t W^{(\vep)}_t \approx \mathcal{C}[W^{(\vep)}_t]$ where $\mathcal{C}$ denotes the appropriate phonon Boltzmann collision 
operator.  The explicit form is straightforward to compute 
using (\ref{eq:Ctotform2}) and the fact that 
$\delta(\omega_0+\omega_1-\omega_2-\omega_3)\left(W_0^{-1}+W_1^{-1}- W_2^{-1} -W_3^{-1}\right)=0$,
for any of the functions $W=\Weql$ in  (\ref{eq:defWeql}).
We obtain the linearized Boltzmann equation $\partial_t h_t = - \mathcal{L} h_t$ where $\mathcal{L}$ is the operator
for which $\mathcal{L} h= L W^{-2} h$ with $W^{-1}$ denoting multiplication by the function $\beta \omega(k)$ and 
\begin{align}\label{eq:defL}
 &
 (L f)(k_0) =
 \int_{\T^3}\! \rmd k\,
\delta(k_0+k_1-k_2-k_3) \delta(\omega_0+\omega_1-\omega_2-\omega_3)
 \nonumber \\ & \quad \times
\prod_{\ell=0}^3 \left[G(\omega_\ell) W_\ell\right]
 \left(f_0+f_1-f_2-f_3\right)
\, ,
\end{align}
with $W_\ell=1/(\beta \omega(k_\ell))$ and $f_\ell = f(k_\ell)$.

The operator $L$ is self-adjoint, even positive, on the Hilbert space $L^2(\T)$:
if $f\in L^2(\T)$ is any sufficiently regular function, then by the symmetry properties of the integrand,
the scalar product between $f$ and $Lf$ satisfies
\begin{align}\label{eq:Lispos}
 & \braket{f}{L f}
 = \frac{1}{4} \int_{\T^{\set{0,1,2,3}}}\! \rmd k\,
\delta(k_0+k_1-k_2-k_3) \delta(\omega_0+\omega_1-\omega_2-\omega_3)
 \nonumber \\ & \quad \times
\prod_{\ell=0}^3 \left[G(\omega_\ell) W_\ell\right]
 \times \left|f_0+f_1-f_2-f_3\right|^2 \ge 0
\, .
\end{align}
In fact, the positivity of the operator $L$ is closely related to the maximization of entropy at the steady state.
Namely,
by differentiating the entropy production (\ref{eq:derHtheorem}) twice, we find
\begin{align}%\label{eq:derHtheorem}
& \partial_t \partial_\vep^2 S[W^{(\vep)}_t]|_{\vep=0} = 2 \braket{f}{Lf}
\, , \quad \text{with } f= W^{-2} h_t
 \, ,
\end{align}
where we have used the above mentioned property that the integrand vanishes whenever either of the two factors
involving differences of $W^{-1}$ is left undifferentiated.

Since $L$ is positive operator, any operator of the form $A^\dagger L A$, where $A$ is a bounded operator and $A^\dagger$ denotes its adjoint, is also positive.  
Therefore, the operator $\tilde{L}= W^{-1} L W^{-1}$ is positive on $L^2(\T)$, and 
thus the solution of the evolution equation $\partial_t \tilde{h}_t = - \tilde{L} \tilde{h}_t$
can be written as $\tilde{h}_t=\rme^{-t \tilde{L}} \tilde{h}_0$ where each $\rme^{-t \tilde{L}}$
is a contraction operator on $L^2(\T)$. 

These properties indicate that it is more natural to study the perturbations in terms of
$\tilde{h}_t$ instead of $h_t$: any solution to 
$\partial_t h_t = - \mathcal{L} h_t$ provides a solution to $\partial_t \tilde{h}_t = - \tilde{L} \tilde{h}_t$
by setting $\tilde{h}_t = W^{-1} h_t$ and vice versa.  Therefore, the Boltzmann equation
linearized around a steady state $W$ has a solution
\begin{align}\label{eq:htsol}
 h_t = W \rme^{-t \tilde{L}} W^{-1} h_0\, ,
\end{align}
for every initial perturbation $h_0$ for which $W^{-1} h_0\in L^2(\T)$.

As mentioned above, 
the Green--Kubo correlation function concerns the case with 
$h_0(k) = W(k)^2 \sin (2\pi k)$ and $W(k)= \beta^{-1} \omega(k)^{-1}$.  
Thus $W^{-1} h_0$ is equal to 
\begin{align}\label{eq:defht0}
\tilde{h}_0(k)=\beta^{-1} \frac{\sin (2\pi k)}{\omega(k)}
\, , 
\end{align}
%$\tilde{h}_0(k)=\beta^{-1} \sin (2\pi k)/\omega(k)$ 
which is a bounded
function on the torus; for FPU-like models with $\omega(k)=O(|k|)$,
the function $h_0$ is not continuous at $k=0$ but its left and right limits exist and are finite.
Hence, we can use the solution in (\ref{eq:htsol}) and conclude that
the Green--Kubo correlation function at the steady state with temperature $\beta^{-1}$ 
has a kinetic theory approximation
\begin{align}\label{eq:Ctkin}
 C(t;\beta) \approx  \ombasic^4 \delta^2 \braket{\phi_0}{h_t}
 =  \ombasic^4\delta^2 \int_\T \! \rmd k \, \phi_0(k) W(k) \left(\rme^{-t \tilde{L}} \tilde{h}_0\right)(k)
 = \ombasic^4 \delta^2 \braket{\tilde{h}_0}{\rme^{-t \tilde{L}} \tilde{h}_0}\, .
\end{align}

The right hand side is equal to the $L^2$-norm 
$\norm{\ombasic^2\delta \rme^{-\frac{1}{2} t \tilde{L}} \tilde{h}_0}^2$, in particular, it is always positive.
The question about the prediction of kinetic theory for the thermal conductivity at a certain equilibrium
state hence boils down to the decay of the above norm under the semigroup $\rme^{-t \tilde{L}}$.
As we will show next, the two Boltzmann equations discussed above prove that both integrable and
non-integrable decay can occur in the kinetic theory of phonons.

\subsection{Kinetic theory prediction for thermal conductivity in chains with anharmonic pinning}
\label{sec:kappapinning}

Suppose the positive operator $\tilde{L}$ has a spectral gap of size $\delta_0>0$ above its zero eigenvalue
and $\tilde{h}_0$ is orthogonal to its eigenspace of zero.  Then spectral theory implies a bound
$\norm{\rme^{-\frac{1}{2} t \tilde{L}} \tilde{h}_0}^2\le 
\norm{\tilde{h}_0}^2\rme^{-t \delta_0}$
which is integrable over $t$.  Hence, in this case the kinetic theory 
prediction is always a finite conductivity
and its leading behaviour in $\lambda_4$ 
can be computed using the kinetic approximation in (\ref{eq:GK1}).  This yields 
\begin{align}\label{eq:kappafinite}
 & \kappa(\beta^{-1}) \approx \beta^2 \ombasic^4 \delta^2 \int_0^\infty\!\rmd r\, 
 \braket{\tilde{h}_0}{\rme^{-r \tilde{L}} \tilde{h}_0}
 = \beta^2 \ombasic^4 \delta^2 \lim_{\vep\to 0^+} \int_0^\infty\!\rmd r\, 
 \braket{\tilde{h}_0}{\rme^{-r (\vep+\tilde{L})} \tilde{h}_0}
  \nonumber \\ & \quad 
 = \lim_{\vep\to 0^+} \beta^2 \ombasic^4 \delta^2 
 \braket{\tilde{h}_0}{(\vep+\tilde{L})^{-1} \tilde{h}_0}
\, ,
\end{align}
where the middle equality is a consequence of dominated convergence theorem.

We have added the regulator $\vep>0$ here to make the operator $\tilde{L}$ invertible
on the whole Hilbert space.  Then the 
scalar product $\braket{\tilde{h}_0}{(\vep+\tilde{L})^{-1} \tilde{h}_0}$ is
finite even if $\tilde{h}_0$ is not orthogonal to the zero subspace of $\tilde{L}$.
This has the benefit that one can use the full operator $\tilde{L}$, instead
of its restriction to the orthocomplement of the zero eigenspace,
and study the 
expectation of the resolvent,
\begin{align}
 R(\vep) = \braket{\tilde{h}_0}{(\vep+\tilde{L})^{-1} \tilde{h}_0}\, , \quad \vep>0\,,
\end{align}
instead of the asymptotic behaviour of the semigroup.

The above result implies that, should $\lim_{\vep\to 0^+} R(\vep)=\infty$, then the kinetic prediction for 
the thermal conductivity at the corresponding steady state is also infinite (note that both $\tilde{L}$ and $\tilde{h}_0$
depend on the choice of the steady state).  In particular, this happens if $\tilde{h}_0$ is not
orthogonal to the zero eigenspace of $\tilde{L}$: 
otherwise, 
we have $R(\vep)\ge \norm{P_0\tilde{h}_0}^2/\vep\to \infty$, as $\vep\to 0$, where $P_0$ denotes the 
orthogonal projection to the zero eigenspace.  Hence, the orthogonality condition mentioned in the beginning of this subsection
is necessary for a finite prediction from kinetic theory.

Let $\tilde{\mu}(\rmd \alpha)$ denote the spectral measure obtained from the spectral
decomposition of $\tilde{L}$ with respect to the vector $\tilde{h}_0$.  Then $\tilde{\mu}$ is a positive Borel measure on the
real axis whose support lies on the spectrum of $\tilde{L}$,
on $\sigma(\tilde{L})\subset \R_+$, and it is characterized by the condition 
\begin{align}
\braket{\tilde{h}_0}{f(\tilde{L}) \tilde{h}_0} = \int \tilde{\mu}(\rmd \alpha) f(\alpha)\, ,
\end{align}
which holds, in particular, for every continuous real function $f$.
Thus $\norm{\tilde{h}_0}^2=\int \tilde{\mu}(\rmd \alpha)$, and 
\begin{align}\label{eq:defRvep}
 R(\vep) = \int_{\sigma(\tilde{L})}\! \tilde{\mu}(\rmd \alpha)\, \frac{1}{\vep+\alpha}\,.
\end{align}

If $\tilde{h}_0=0$, we have $R(\vep)=0$ for all $\vep$, and thus in this case the kinetic theory prediction would be zero conductivity.
If $\tilde{h}_0\ne 0$, such as for the above discussed particle chains, 
we can normalize the measure $\tilde{\mu}$ into a probability measure
by dividing it by $\norm{\tilde{h}_0}^2$.  In addition,
the second derivative of the function $\alpha\mapsto (\vep{+}\alpha)^{-1}$ is positive, and thus it is convex.  Therefore, 
we can apply Jensen's inequality in (\ref{eq:defRvep}) to conclude that
\begin{align}%\label{eq:defRvep}
 R(\vep) =\norm{\tilde{h}_0}^2 \int_{\sigma(\tilde{L})}\frac{\tilde{\mu}(\rmd \alpha)}{\norm{\tilde{h}_0}^2} \frac{1}{\vep+\alpha}
 \ge \frac{\braket{\tilde{h}_0}{\tilde{h}_0}^2}{\vep\braket{\tilde{h}_0}{\tilde{h}_0} + \braket{\tilde{h}_0}{\tilde{L}\tilde{h}_0}}\,.
\end{align}
The right hand side converges to $\braket{\tilde{h}_0}{\tilde{h}_0}^2/\braket{\tilde{h}_0}{\tilde{L}\tilde{h}_0}$
as $\vep\to 0$.  Since $\tilde{L}=W^{-1} L W^{-1}$ and $\tilde{h}_0=W \phi_0$, we find that
\begin{align}\label{eq:Jensenbound}
\beta^2 \ombasic^4 \delta^2 \frac{\braket{W \phi_0}{W \phi_0}^2}{\braket{\phi_0}{L \phi_0}}
\end{align}
forms a \defem{lower bound} for the kinetic theory prediction of thermal conductivity in the particle chains.

By (\ref{eq:Lispos}), $\psi$ can be a zero eigenvector of $\tilde{L}$ if and only if it belongs to $L^2(\T)$
and $f=W^{-1} \psi$ satisfies (\ref{eq:steadyeq}), i.e., it is a collisional invariant.  We have argued (and even proven for $\delta=1/2$) that 
all collisional invariants are linear combinations of $\omega$ and $1$.  This implies that 
the zero subspace of $\tilde{L}$ is spanned by $1$ and $W$.
For the particle chains, $\tilde{h}_0(k)=W(k) \phi_0(k)=\omega'(k)/(2\pi \delta\ombasic^2 \beta)$ where $\omega'$ denotes the derivative of $\omega$.
Therefore, by periodicity, $\int_\T \rmd k\, \tilde{h}_0(k) (a W(k)+b)=0$ for all $a,b\in \C$, and thus $\tilde{h}_0$ is indeed
orthogonal to the proposed zero subspace of $\tilde{L}$.  

Let us also sketch a possible proof for the spectral gap of $\tilde{L}$.
Consider an arbitrary $\psi$ and denote $f=W^{-1} \psi$.
We
begin from the integral representation (\ref{eq:Lispos}) for 
$\braket{f}{Lf}=\braket{\psi}{\tilde{L}\psi}$.  Let us mollify the singularities in the integrand by choosing a suitable
regularizing function $\Phi:\T^4 \to [0,1]$ and then using the bound $1\ge \Phi(k)$ inside the integrand.
The purpose of $\Phi$ is to regularize the collision cross section of the model,
so that in the remaining integral one can use the explicit solution and define an operator $L'$
such that $\braket{f}{Lf}\ge \braket{f}{L'f}$ and $L'=V-K$ where $V$ is a positive multiplication operator and
$K$ is a self-adjoint integral operator.  If we can then tune $\Phi$ so that $V$ and $1/V$ are both bounded 
and $B=V^{-1/2}K V^{-1/2}$ is a compact operator, it follows that $1-B$ has a spectral gap and its zero subspace consists of those
functions $\phi$ for which $V^{-1/2}\phi$ is a collisional invariant.  Then for every $f$ which is orthogonal 
to the collisional invariants, we have 
$\braket{f}{Lf}\ge \braket{f}{L'f}\ge \delta' \norm{f}^2$ for some $\delta'>0$ independent of $f$.
This estimate would then prove a gap also for 
$\tilde{L}$ in the pinned case: then $W$ and $1/W$ are both bounded functions and there is $\delta_0>0$ such that 
$\braket{\psi}{\tilde{L}\psi}\ge \delta_0 \norm{\psi}^2$ for every 
$\psi$ for which $W^{-1} \psi$ is orthogonal to the collisional invariants.

The above procedure is a variant of the standard argument used to prove a gap in kinetic theory when the ``relaxation time''
determined by the multiplication operator $V$ is bounded from above and below (see the next section for discussion about the
relaxation time approximation).  The problem about using the standard argument directly for
the linearized operator of the pinned chains lies in the non-integrable singularity related to the Jacobian of the change of variables
which resolves the energy constraint.  This singularity is the same one which makes the total collision cross section---obtained
by neglecting all terms which contain $W_0$ in (\ref{eq:resolvedonsiteBE})---infinite: since in the integrand
$p_3=p_1-y$ when $p_2=p_0+y$, the integrand has a nonintegrable singularity of strength at least $|y|^{-1}$
at $p_2=p_0$.  Thus both the total collision cross section and the relaxation time function $V$ are formally infinite for all $k_0$.
The sign however is positive, and thus by reducing the strength of the collisions one can aim at approximating the linearized operator from
below by an operator of the ``standard form'' with a finite relaxation time function.  However, 
let us not try to complete the argument here but instead focus on its implications on the thermal conductivity.

To summarize, the above argument strongly indicates that the kinetic theory prediction for the thermal conductivity 
at temperature $\beta^{-1}$ for
chains with anharmonic pinning is a finite non-zero value, and for small $\lambda_4$ we should have
\begin{align}%\label{eq:defRvep}
  & \kappa(\beta^{-1}) \approx \beta^2 \ombasic^4 \delta^2 
  \braket{\tilde{h}_0}{\tilde{L}^{-1} \tilde{h}_0}
\, ,
\end{align}
where ``$\tilde{L}^{-1}\tilde{h}_0$''
denotes the unique $\phi \in L^2(\T^d)$ for which $\tilde{L}\phi =\tilde{h}_0$ (such  $\phi$ can be found if $\tilde{L}$ has 
a spectral gap and $\tilde{h}_0$ is orthogonal to the zero eigenspace of $\tilde{L}$, as we have argued above).
Inserting $W(k)=(\beta \omega(k))^{-1}$ and using the definitions of $\tilde{L}$ and $\tilde{h}_0$ this can be simplified to 
\begin{align}%\label{eq:defRvep}
  & \kappa(\beta^{-1}) \approx \beta^{-2} \ombasic^4 \delta^2 
  \braket{\omega^{-2} \phi_0}{L^{-1} (\omega^{-2} \phi_0)}
\, ,
\end{align}
where the operator $L$ is explicitly given by 
\begin{align}\label{eq:defLonsite}
 &
 (L \psi)(k_0) = \beta^{-4} \lambda_4^2 \frac{9\pi}{4}
 \int_{\T^3}\! \rmd k\,
\delta(k_0+k_1-k_2-k_3) \delta(\omega_0+\omega_1-\omega_2-\omega_3)
 \nonumber \\ & \quad \times
\prod_{\ell=0}^3 \omega_\ell^{-2}
 \left(\psi_0+\psi_1-\psi_2-\psi_3\right)
\, .
\end{align}

Thus the dependence on the temperature and coupling factorizes:
\begin{align}%\label{eq:defRvep}
  & \kappa(\beta^{-1}) \approx \beta^{2} \lambda_4^{-2} \ombasic^9 C(\delta)
\, ,
\end{align}
where the constant $C(\delta)$ is a function of the harmonic pinning parameter $\delta$ only, 
\begin{align}%\label{eq:defRvep}
& C(\delta) = \delta^{3} \frac{8}{9} 
  \braket{\nu^{-2} \phi_0}{L_0^{-1} (\nu^{-2} \phi_0)}\, ,\quad\text{with}\\
& 
 (L_0 \psi)(k_0) = 
 \int_{-\pi}^\pi\! \frac{\rmd p_2}{2\pi}\, \left|\frac{\sin p_1}{\nu_1}-\frac{\sin p_3}{\nu_3}\right|^{-1}
\prod_{\ell=0}^3 \frac{1}{\nu_\ell^{2}}
 \left(\psi_0+\psi_1-\psi_2-\psi_3\right)\, ,
\end{align}
where $\psi_\ell=\psi(p_\ell/(2\pi))$, 
$p_0=2\pi k_0$, $p_3=p_1+p_0-p_2$, and $p_1=h(p_0,p_2;\delta)$, as defined in (\ref{eq:defhnonptonsite}).
In addition, we have used here $\nu$ to denote the normalized dispersion relation with $\ombasic=1$,
that is, in the above formulae $\nu(k)=\sqrt{1-2\delta \cos (2\pi k)}$ and 
$\nu_\ell = \nu(p_\ell/(2\pi))$.
The form is amenable for numerical inversion of the operator $L_0$, by choosing a suitable orthonormal 
basis for the subspace of $L^2(\T)$ which consists of vectors orthogonal to $1$ and $\omega$.

The Jensen inequality lower bound given in (\ref{eq:Jensenbound}) implies that
\begin{align}
& C(\delta) \ge \delta^{3} \frac{8}{9} \frac{\braket{\nu^{-1}\phi_0}{\nu^{-1}\phi_0}^2}{\braket{\phi_0}{L_0 \phi_0}}\, .
 \end{align}
Using the symmetrized form in (\ref{eq:Lispos}) for $\braket{\phi_0}{L_0 \phi_0}$ and changing the integration variable from $k_0$
to $p_0=2\pi k_0$ then yields
 \begin{align} \label{eq:smalldJensen}
& \delta^{-3} C(\delta) \ge \frac{32}{9} \left(\int_{-\pi}^\pi \!\rmd p_0\, \frac{\sin^2 p_0}{\nu_0^2}\right)^2
 \nonumber \\ & \quad \times
\biggl(\int_{-\pi}^\pi \!\rmd p_0\int_{-\pi}^\pi \!\rmd p_2\, \left|\frac{\sin p_1}{\nu_1}-\frac{\sin p_3}{\nu_3}\right|^{-1}
\prod_{\ell=0}^3 \frac{1}{\nu_\ell^{2}}
 \left|
 \psi_0+\psi_1-\psi_2-\psi_3\right|^2
\biggr)^{-1}\, ,
 \end{align}
where $\psi_\ell=\sin p_\ell$.  When $\delta\to 0$, we have $\nu_\ell\to 1$, $p_1\to \pi -p_0$, and $p_1\to \pi -p_2$
in the above.  The remaining integrals can be computed explicitly, and the limit of the right hand side 
is found to be equal to $\pi^2/36\approx 0.274$.  The numerical inversion of the full operator 
in the $\delta\to 0$ limit in \cite{ALS06} resulted in the value $0.2756$ which is very close to the 
above Jensen bound.  In addition, evaluation of the right hand side of (\ref{eq:smalldJensen})
by numerical integration shows that it depends only weakly on $\delta$, decreasing to $0.2$ at $\delta=0.3$.
However, the bound becomes ineffective for larger values of $\delta$, going to zero as $\delta\to \frac{1}{2}$.

\begin{figure}[t]
  \centering
  \includegraphics*[width=0.7\textwidth]{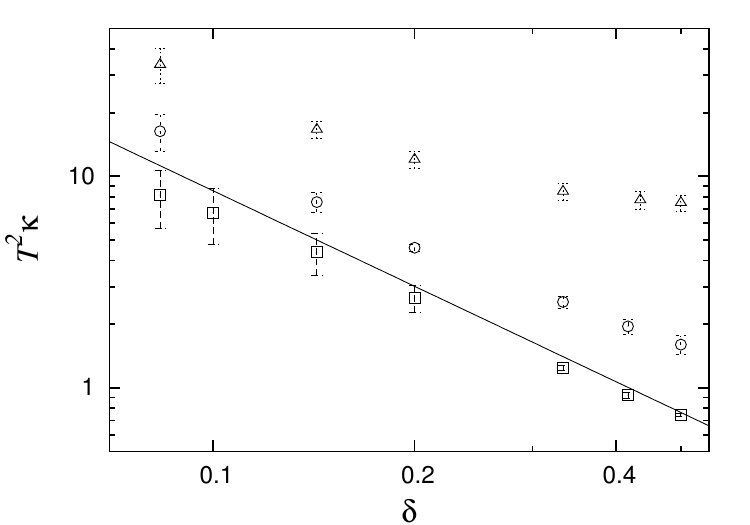}
  \caption{Measured $T^2\kappa(T)$ compared against the
  kinetic theory computation for small $\delta$ and $T$ (straight line).
  The data points are for $T=0.1\ (\Box)$, $0.4\ (\circ)$, and $4
  (\bigtriangleup)$. (Reprinted from \cite{ALS06})}
  \label{fig:dDep}
\end{figure}

Therefore, the formula $0.274\, \delta^3  \ombasic^9 \beta^{2} \lambda_4^{-2}$
provides a fairly good approximation for the lower bound in (\ref{eq:Jensenbound})
for the kinetic theory prediction of the thermal conductivity, at least for small enough $\delta$.
This approximation 
was compared in \cite{ALS06} to the thermal conductivity measured in numerical simulations of large finite chains with
boundary thermostats.
The numerical simulations where performed with $\ombasic=\delta^{-1/2}$ and $\lambda_3=0$,
and instead of $\lambda_4 \to 0$ with fixed $\beta$ one considers $\beta\to \infty$ with fixed $\lambda_4$.  The two limits can be connected
by a straightforward scaling argument which allows to compare the kinetic prediction with the conductivity observed
in the simulations (see \cite[Section 2]{ALS06} for details).
As mentioned above, the Jensen inequality lower bound given in (\ref{eq:Jensenbound})
is a good approximation of the numerically inverted value for small $\delta$.  In Fig.\ \ref{fig:dDep} we have given the
comparison between the values from the numerical particle simulations and the Jensen bound.  The agreement is surprisingly good
and seems to indicate that for this model the kinetic theory gives good description of the 
dominant effects affecting thermal conduction
in the pinned anharmonic chains.

\subsection{Anomalous energy conduction in the kinetic theory of FPU chains}
\label{sec:CtFPU}

The kinetic approximation to the Green--Kubo correlation function in the FPU chains has been studied
rigorously in \cite{ls07}.  As mentioned above, the FPU chains have $\delta=\frac{1}{2}$, and 
in \cite{ls07} the frequency normalization was chosen as $1/\sqrt{2}$ since then $\omega(k) = |\sin(\pi k)|$ 
has no prefactors.  For easy comparison, let us make this choice also in this subsection for the FPU chains: 
if needed, the scale $\ombasic$ can be reintroduced in the results just as was done in the previous subsection.

With this choice and using (\ref{eq:defL}), the Boltzmann collision operator linearized around a steady state $W(k)= \beta^{-1} \omega(k)^{-1}$ becomes
\begin{align}\label{eq:defLFPU}
 &
 (L f)(k_0) = 9 \pi \lambda_4^2 (2/\beta)^4
 \int_{\T^3}\! \rmd k\,
\delta(k_0+k_1-k_2-k_3) \delta(\omega_0+\omega_1-\omega_2-\omega_3)
 \nonumber \\ & \quad \times
 \left(f_0+f_1-f_2-f_3\right)
\, ,
\end{align}
and it is related to the kinetic theory prediction of the Green--Kubo correlation function by 
\begin{align}%\label{eq:Ctkin}
 C(t;\beta) \approx  \frac{1}{2^4} \braket{\tilde{h}_0}{\rme^{-t \tilde{L}} \tilde{h}_0}\, ,
\end{align}
where $\tilde{L}=\beta^2 \omega L \omega$ and $\tilde{h}_0(k)=2 \cos(\pi k)/\beta$ for $k\in [0,1)$.  Choosing
a normalization somewhat different from the previous subsection, we may rewrite this
in a dimensionless form by defining
$\psi(k)= \frac{1}{2}\cos(\pi k)$, $0\le k<1$, and $L_0 = \frac{1}{9} \lambda_4^{-2} (\beta/2)^4 \omega L \omega$.
This yields
\begin{align}%\label{eq:Ctkin}
 C(t;\beta) \approx  \frac{1}{\beta^2} \braket{\psi}{\rme^{-t c \tilde{L}_0} \psi}\, ,
\end{align}
where the constant $c=12^2\lambda_4^2 \beta^{-2} \pi^{-1}$ and $\tilde{L}_0=\omega L_0 \omega$.
In \cite{ls07}, this was called the kinetic conjecture and the above form coincides with the one given in 
\cite[Eq.~(1.18)]{ls07} after a change of variables from $k\in [0,1)$ to $p\in [0,2\pi)$.

Explicitly, the operator $L_0$ can be written for $p\in I=[0,2 \pi)$ as
\begin{align}%\label{eq:defLFPU}
 &
 (L_0 f)(p_0/(2\pi)) = 
 \int_{I^3}\! \rmd p\,
\delta(p_0+p_1-k_2-k_3) \delta(\omega_0+\omega_1-\omega_2-\omega_3)
 \nonumber \\ & \quad \times
 \left(f_0+f_1-f_2-f_3\right)
\, .
\end{align}
Now the constraints can be explicitly integrated using the results mentioned in Sec.~\ref{sec:FPUconst},
yielding
\begin{align}%\label{eq:defLFPU}
 & L_0 = V - A\, ,\quad \text{with}\\
 & (V\psi)(k)= V(2\pi k)\psi(k)\qand (A\psi)(k) = \int_0^1 \! \rmd k' 2\pi K(2\pi k,2\pi k') \psi(k')\, ,
\end{align}
where
\begin{align}%\label{eq:defLFPU}
 & V(p) = \int_0^{2\pi}\!\rmd p' K_2(p,p')\, ,\quad
 K(p,p') = 2 K_2(p,p')- K_1(p,p')\, ,\quad \text{with}\\
 & K_1(p,p')  =
  4\frac{\1(F_-(p,p')>0)}{\sqrt{F_-(p,p')}} \qand
  K_2(p,p') =  \frac{2}{\sqrt{F_+(p,p')}}
\, .
\end{align}
In this formula, the multiplication operator arises from the ``$f_0$''-term and
the $K_1$ term in the integral kernel $K$ from the ``$f_1$''-term.
By symmetry, the contributions from the ``$f_2$''-  and ``$f_3$''-terms 
are equal, each contributing one $K_2$-term to the integral kernel.
It probably comes as no surprise that the precise analysis of the 
semigroup generated by $\tilde{L}_0$ gets rather technical.  
However, 
this has been done in \cite{ls07}, and let us only repeat the main conclusions
from the analysis here.

Unlike for the onsite anharmonic perturbation, the relaxation time for the FPU models
is finite, and as the above formula shows, the operator $\tilde{L}_0$
can be written in the standard form $\tilde{V} - \tilde{A}$
where $\tilde{V}=\omega^2 V$ is a multiplication operator and $\tilde{A}=\omega A\omega$
is an integral operator.  It is proven in \cite[Lemma 4.1]{ls07} that
the function $\tilde{V}(k)$ is continuous and can be bounded from above and below by $|\sin(\pi k)|^{5/3}$.
In particular $\tilde{V}(0)=0$ and, consequently, the operator $\tilde{L}_0$
has no spectral gap.

In kinetic theory, it is a common practice to use the \defem{relaxation time
approximation} to approximate the linearized Boltzmann evolution.
In our case, this amounts to dropping the operator $\tilde{A}$, i.e., approximating
\begin{align}%\label{eq:}
\braket{\psi}{\smash{\rme^{-t \tilde{L}_0}} \psi } \approx
\braket{\psi}{\smash{\rme^{-t \tilde{V}}} \psi } 
\, .
\end{align}
Since $\psi(0)=1$, the decay of the relaxation time
approximation is now entirely determined by the values of the potential near zero.
The above bounds imply that $\tilde{V}(k) = O(|k|^{5/3})$, and thus the 
relaxation time approximation predicts $C(t;\beta) = \order{t^{-3/5}}$ for
large $t$.  (To our knowledge, this decay of the relaxation time approximation was 
first as derived in \cite{perev03}.)

However, the contribution arising from adding the integral operator $\tilde{A}$
is also singular, and more careful study is required to conclude that the relaxation 
time prediction continues to hold for the full semigroup.
Fortunately, the above straightforward estimate gives the correct decay speed: it
was shown in \cite{ls07} that the resolvent of the semigroup satisfies
\begin{align}\label{eq:resolv1a}
& \bigbraket{\psi}{\frac{1}{\vep +\tilde{L}_0} \psi} 
= \bigbraket{\psi}{\frac{1}{\vep +\tilde{V}} \psi}  + 
\bigbraket{\psi}{\frac{1}{\vep +\tilde{V}}\tilde{A} \frac{1}{\vep +\tilde{V}} \psi}  
% \nonumber \\ & \qquad 
+ \bigbraket{\psi}{\frac{1}{\vep +\tilde{V}}\tilde{A} \frac{1}{\vep +\tilde{L}}\tilde{A}
  \frac{1}{\vep +\tilde{V}} \psi} ,
\end{align}
where the first term is identical to the relaxation time approximation, and behaves
as $\vep^{-2/5}$ for small $\vep>0$.  The second and third term are
$\order{\vep^{-1/5-\vep}}$ for any $\vep>0$.  Although also this
second contribution is divergent, the first term is dominant, and thus we
confirm the prediction of the relaxation time approximation in this particular
case.

Even the exact asymptotics can be found for this particular choice.  Applying \cite[Corollary 2.6]{ls07}
we find that the kinetic theory predicts for times with $t (\lambda_4/\beta)^2\gg 1$, 
and for sufficiently small couplings $\lambda_3$ and $\lambda_4$,
\begin{align}
 C(t;\beta) \approx C_0 \beta^{-\frac{4}{5}} (\lambda_4^2 t)^{-\frac{3}{5}} \, .
\end{align}
Here $C_0=c_0(\pi 12^{-2})^{3/5}/(2 \pi \Gamma(2/5))$ is an explicit numerical constant.
Evaluation of the Gamma-function and the integrals defining the constant ``$c_0$'' in (6.14) and (4.7)
of \cite{ls07} yields a numerical approximation $C_0 \approx 0.00386$.

Of course, for much longer than kinetic times, the terms neglected in the derivation of the Boltzmann equation might 
become important and alter the asymptotic decay.
The above results also imply that, on the kinetic time scale, the energy spread is
superdiffusive: the quadratic energy spread observable
discussed in Sec.~\ref{sec:GKformula} should then be increasing as $S(t) = O(t^{7/5})$.

The energy spread has been analysed in more detail in \cite{MM15}.  The authors study the time evolution
of \defem{inhomogeneous} perturbations around a given thermal equilibrium state.  This results in an evolution
equation which corresponds to the phonon Boltzmann evolution where the nonlinear collision operator has been replaced by the 
above linearized operator.   Explicitly, the perturbation $f(t,x,k)$, defined via $W=(1+f)/(\beta \omega)$, evolves then by 
\begin{align}\label{eq:inhomogBoltzFPU}
 \partial_t f(t,x,k) + \frac{1}{2\pi}\omega'(k) \partial_x f(t,x,k) = (\tilde{L} f(t,x,\cdot))(k)\, .
\end{align}
The result concerns $L^2$-integrable initial data varying at a scale $\vep^{-1}$, with $\vep$ small.
It is shown that for sufficiently long times, the solution then first thermalizes in the $k$-variable, 
which by the definition of $f$ implies that it becomes independent of $k$.
Diffusive relaxation in the spatial variable would then mean that at a time scale $\vep^{-2}$
the perturbation follows the heat equation $\partial_t f + \kappa (-\Delta) f=0$, with $\kappa>0$.
However, this does not occur here: instead, it is shown that 
at the time scale $\vep^{-8/5}$ the perturbation satisfies a fractional diffusion equation 
$\partial_t f + \kappa (-\Delta)^{4/5} f=0$, with $\kappa>0$.  

This corresponds to a superdiffusive relaxation of the initial perturbation.  Moreover, the
fractional diffusion spreads local perturbations at time $t$ only up to distances 
$t^{5/8}$.  This is in apparent contradiction with the earlier claim that $S(t) = O(t^{7/5})$
which would indicate that the spatial spread occurs at a speed $O(t^{7/10})$, i.e., faster
than predicted by the fractional diffusion equation.

The resolution lies in the tail behaviour of solutions to the 
fractional diffusion equation.  Let us conclude the section with a somewhat heuristic argument which would 
explain the above results.
By using Fourier-transform and a simple scaling argument,
one finds that the solution to $\partial_t f + \kappa (-\Delta)^{4/5} f=0$, with an initial data
$f_0$, is given by a convolution of $f_0$ with an integral kernel $K_t$ which satisfies
a scaling relation $K_t(x)=t^{-p} F(x/t^p)$ where $p=5/8$.  Unlike the Gaussian heat kernel in (\ref{eq:gausssol}), 
the function $F$ is merely polynomially decreasing, with $F(y)$ decaying as $y^{-13/5}$ for large $|y|$
(the power $13/5$ is obtained by evaluating $1/p+1$).
Thus $\int\! \rmd y\, y^2 F(y)=\infty$, and 
the quadratic spread from fractional diffusion becomes immediately infinite, even if it is initially finite.

However, on the microscopic scale, the velocities of the ballistic harmonic evolution are bounded from above, and 
this will eventually cut off the above
decay, and change it from the above powerlaw decay to an exponentially fast decay at spatial 
distances $O(t)$ from the source.  Therefore, we would expect that the true microscopic distribution 
of an initially local perturbation at a large time $t$ is $O(t^{-p})$ for distances $|x|=O(t^{p})$,
it is $O(t |x|^{-2-3/5})$ for $|x|$ between $O(t^{p})$ and $O(t)$, and it becomes exponentially decreasing for
distances larger than $O(t)$.  For such functions, the value of $S(t)$ is entirely dominated by the 
midscale powerlaw tail which yields a term $O(t^{7/5})$, just as we obtained from the kinetic
prediction for the Green--Kubo correlation function using the 
linearized Boltzmann equation.

\section{Concluding remarks}
\label{sec:conclusions}

As the last two explicit examples show, kinetic theory of phonons is capable of uncovering detailed
information about the decay of time-correlations and, via the Green--Kubo formula, about the thermal conductivities
of classical particle chains.  This is somewhat surprising, considering the various mathematical problems and uncertainties
discovered along the way to the phonon Boltzmann equation and even in its analysis.  However, 
the agreement between the kinetic prediction and numerical simulations for 
thermal conductivities in chains with anharmonic pinning, and the discovery of anomalous energy
transport by fractional Brownian motion from the linearized Boltzmann equation in the FPU-$\beta$ chains,
present a strong case in favour of looking for further applications of the phonon Boltzmann equations, even
in the somewhat degenerate one-dimensional case.  After all, it is at present one of the very few general tools which allow
computing the dependence of the thermal conduction properties on the parameters of the microscopic evolution
directly, without introduction of additional fitting parameters.

We have also seen that some care is needed in the application of the phonon Boltzmann equation.  
Most importantly, the equation is closely tied to the scales on which the free streaming of phonons \defem{begins}
to alter its character due to the collisions.  Hence, it describes the evolution up to the kinetic time-scale only,
and it is possible that further changes are found at larger time-scales.
Nevertheless, since the H-theorem implies that solutions to the kinetic equation push the system towards thermal equilibrium states,
drastic changes in the character of the evolution should be the exception, not the rule.

The evaluation of the decay of Green--Kubo correlation functions is one of the robust applications of kinetic theory,
and the thermal conductivity obtained by the perturbation procedure recalled here should in general yield its leading
behaviour in the limit of weak perturbations.  However, as a warning about the standard procedure, let us stress that 
the commonly used relaxation time approximation of the linearized collision operator is \defem{only} an order-of-magnitude estimate
of the real kinetic prediction which involves inverting the full linearized operator.  In the pinned case, we found that the
relaxation time approximation predicts zero thermal conductivity.  This turned out to be misleading since already the straightforward Jensen bound
for the full inverse proves that the kinetic prediction for the conductivity is non-zero.  In contrast, for the anomalous conduction in the FPU-$\beta$ chain
the relaxation time approximation does capture the correct asymptotic decay of the kinetic prediction.

One major open question in the kinetic theory of phonons, and in fact of nonlinearity perturbed systems in general, is the precise 
manner of handling spatially inhomogeneous perturbations.  The standard Boltzmann transport term, appearing on the left hand side 
of (\ref{eq:inhomogBoltzFPU}), might require adjusting to capture all effects relevant to the transport.  One such example often 
found in the literature is an addition of a Vlasov-type term.  For which systems, under which time-scales, and for which initial data 
such corrections are necessary, remains unresolved at the moment.

One benefit from a better understanding of the behaviour of
inhomogeneous perturbations could be a first-principles derivation of fluctuating hydrodynamics
for these systems, including the precise dependence of its parameters on the microscopic evolution, in the limit of weak couplings.
The application of fluctuating hydrodynamics to the transport in one-dimensional particle chains
has been discussed in \cite{spohn14} and reviewed 
in \cite{spohn15}.
It appears to be the first model which is able to describe the anomalous transport in one-dimensional particle chains
fully in agreement with computer simulations of the spread of localized perturbations.
Connecting it directly to the microscopic dynamics would be a breakthrough in understanding the microscopic origin and 
precise nature of transport in crystalline structures, such as the present particle chains.
Kinetic theory, the phonon Boltzmann equation in particular, could well provide some of the missing steps into this direction.

\subsection*{Acknowledgments}

I am most grateful to Herbert Spohn for his comments and suggestions for improvements.
Most of the discussion here is based on his works and on our joint collaborations.
The related research has been made possible by support from the Academy of Finland, 
and benefited from the support of the project EDNHS ANR-14-CE25-0011 of the French National Research Agency (ANR).  
I am also grateful to Matteo Marcozzi and Alessia Nota for their comments on the manuscript.

% \newcommand{\utildir}[1]{../../texstuff/#1}
% \bibliographystyle{\utildir{abunst_titles}}
% \bibliography{\utildir{myabbr},\utildir{mrabbrev},\utildir{allrefs}}
% \end{document}

\end{document}